\documentclass[traditabstract]{aa}
\usepackage{epsfig}    
\usepackage{lscape}
\usepackage{natbib}
\bibpunct{(}{)}{;}{a}{}{,}    
\usepackage{graphics}
\usepackage{graphicx,graphics}
\usepackage{color} 
\usepackage{times}
\usepackage{amsmath}
\usepackage{amssymb}
\begin{document}

\title {Calibration-based abundances in the interstellar gas of galaxies from slit and IFU spectra} 
      
\author {
         L.~S.~Pilyugin\inst{\ref{ITPA},\ref{MAO}}            \and 
         M.~A.~Lara-L\'{o}pez\inst{\ref{UCM},\ref{IPARCOS}}   \and
         J.~M.~V\'{i}lchez\inst{\ref{IAA}}                    \and
         S.~Duarte Puertas\inst{\ref{Can},\ref{IAA}}          \and
         I.~A.~Zinchenko\inst{\ref{LMU},\ref{MAO}}            \and
         O.~L. Dors Jr.\inst{\ref{UVP}}
}
\institute{Institute of Theoretical Physics and Astronomy, Vilnius University, Sauletekio av. 3, 10257, Vilnius, Lithuania \label{ITPA} \and
  Main Astronomical Observatory, National Academy of Sciences of Ukraine, 27 Akademika Zabolotnoho St, 03680, Kiev, Ukraine \label{MAO} \and
  Departamento de Física de la Tierra y Astrof\'{i}sica, Universidad Complutense de Madrid, E-28040 Madrid, Spain \label{UCM}\and
  Instituto de F\'{i}sica de Part\'{i}culas y del Cosmos IPARCOS, Fac. de Ciencias F\'{i}sicas, Univ. Complutense de Madrid, E-28040, Madrid, Spain \label{IPARCOS}\and
  Instituto de Astrof\'{i}sica de Andaluc\'{i}a, CSIC, Apdo 3004, 18080 Granada, Spain \label{IAA} \and
  D\'epartement de Physique, de G\'enie Physique et d'Optique, Universit\'e Laval, and Centre de Recherche en Astrophysique du Qu\'ebec (CRAQ), Qu\'ebec, QC, G1V 0A6,
  Canada \label{Can} \and 
  Faculty of Physics, Ludwig-Maximilians-Universit\"{a}t, Scheinerstr. 1, 81679 Munich, Germany \label{LMU}  \and
  Universidade do Vale do Para\'{i}ba, Av. Shishima Hifumi, 2911, Cep 12244-000, S\~{a}o Jos\'{e} dos Campos, SP, Brazil \label{UVP}
  }

\abstract{
  In this work we make use of  available Integral Field Unit (IFU) spectroscopy and slit spectra of several nearby galaxies. The pre-existing empirical R and S calibrations for
  abundance determinations are constructed using a sample of H\,{\sc ii} regions with high quality slit spectra. In this paper, we test the applicability of those calibrations
  to the IFU spectra. We estimate the calibration-based abundances obtained using   both the IFU and the slit spectroscopy for eight nearby galaxies. The median values of the slit
  and IFU spectra-based abundances in  bins of 0.1 in fractional radius $R_{g}$ (normalized to the optical radius $R_{25}$) of a galaxy are determined and compared.  We find that
  the IFU and the slit  spectra-based abundances obtained through the R calibration are close to each other, the mean value of the differences of abundances is 0.005 dex and the
  scatter in the differences is 0.037 dex for 38 datapoints. The S calibration can produce  systematically underestimated values of the IFU spectra-based abundances  at high
  metallicities (12 + log(O/H) $\gtrsim$ 8.55), the mean value of the differences  is --0.059 dex for 21 datapoints,   while at lower metallicities the mean value of the differences
  is --0.018 dex and the scatter is 0.045 dex for 36 data points. This evidences that the R calibration produces more consistent abundance estimations between the slit and the IFU
  spectra than the S calibration. We find that the same calibration  can produce close estimations of the abundances using IFU spectra obtained with different spatial resolution and
  different spatial samplings. This is in line with the recent finding that the contribution of the diffuse ionized gas to the large aperture  spectra of H\,{\sc ii} regions  has a
  secondary effect.
}


\titlerunning{Slit and IFU spectra-based abundances}
\authorrunning{Pilyugin et al.}
\maketitle

\section{Introduction}
The investigation of the relations between gas-phase abundance and other characteristics of galaxies is very important in understanding the (chemical) evolution of galaxies.
In addition to the obtained relations being affected by the absolute abundance values (i.e. metallicity scale), which depends on the adopted method for the abundance determinations,
a mandatory condition in constructing such relations is for  the abundances of all the galaxies to be on a unique metallicity scale. This implies that the abundances should be
determined through a single method or through methods producing  abundances on the same metallicity scale. If  abundances obtained through a given method depend on the way of the
spectral measurements (e.g. on a part of the H\,{\sc ii} region measured) then the emission line spectra of H\,{\sc ii} regions in galaxies should be measured in similar way.    

The emission line spectra of  H\,{\sc ii} regions in galaxies available at the present day are measured in different ways. The long slit spectra of H\,{\sc ii}
regions in many galaxies were obtained in numerous investigations \citep[e.g. see compilations in][]{Pilyugin2004,Pilyugin2014,Zurita2021}.
Strictly speaking, the slit spectra are not uniform in the sense that the physical fraction of  H\,{\sc ii} region within the slit can significantly vary for different
reasons: {\it i)} because of variations on the slit widths in different works, {\it ii)} because of variation in  angular sizes of the  H\,{\sc ii} regions (due to the
variation on the physical sizes of H\,{\sc ii} regions and/or distances to galaxies), {\it iii)} because the slit can cross different parts (centre or periphery)
of  H\,{\sc ii} regions. One can adopt (at least, as the first order approximation) that the variations in different parts of  H\,{\sc ii} regions measured in the slit
spectroscopy are random. 

The Integral Field Unit (IFU)  spectroscopy measurements of the galaxy NGC 628 were carried out within the PPAK Integral Field Spectroscopy Nearby Galaxies Survey, PINGS,
\citep{Sanchez2011,Rosales2011}. The 382 fibre bundle has a field of view (FoV) of a hexagonal shape of 74 arcsec × 64 arcsec. Each fibre projects to 2$\farcs$7 in diameter
on the sky, and the fibre-to-fibre distance is 3$\farcs$2, which yields a total filling factor of 0.6. A dithering scheme with three pointings has been used to cover the
complete FoV of the  bundle. The spatial resolution of full width at half maximum (FWHM) is $\sim$2.5 arcsec. The catalog of  H\,{\sc ii} regions in NGC~628 is generated,
and the spectra of detected  H\,{\sc ii} regions  are extracted \citep{Rosales2011}. The angular size of a nearby galaxy usually exceeds  the diameter of the field of
view of the spectrograph. Therefore a number of pointings are observed to form a mosaic, for instance, \citet{Sanchez2011}  obtained 34 pointings for the NGC~628. 

The IFU spectroscopy measurements of a sample of galaxies on the nearby Universe are carried out in the same way (and using the same instrument as in PINGS) within the Calar Alto
Legacy Integral Field Area (CALIFA) survey \citep{Sanchez2012,Sanchez2016,GarciaBenito2015}. Those observations are used to construct a grid of spectra with a spatial sampling of
1$\arcsec$ and/or to detect  H\,{\sc ii} regions and extract their spectra. The recent version of the catalogue of  H\,{\sc ii} regions involves 924
galaxies\footnote{http://ifs.astroscu.unam.mx/CALIFA/HII\_regions/new\_catalogs/} \citep{Espinosa2020}.

The fibre spectra were measured in many galaxies within the Sloan Digital Sky Survey (SDSS, \citet{York2000}). The SDSS spectra are obtained through a 3-arcsec diameter
fibres. In the framework of the SDSS IV programme, the IFU spectroscopy for 10000 galaxies are carried out within the Mapping Nearby Galaxies at Apache Point Observatory
(MaNGA) survey \citep{Bundy2015}. The diameters of the fields of view vary from 12$\arcsec$ (19 fibres) to 32$\arcsec$ (127 fibres).  A final spatial sampling is 0$\farcs$5/pixel,
the point spread function (PSF) of the MaNGA measurements is estimated to have a full width at half maximum of 2.5 arcsec or 5 pixel \citep{Bundy2015,Belfiore2017}.
The wavelength interval covers from 3600 to 10300 {\AA} with the spectral resolution $R$ $\sim$ 2000.

A sample of nearby spiral galaxies (within $\sim$20 Mpc) are measured  within the PHANGS (Physics at High Angular Resolution in Nearby Galaxies) programme,
using the Very Large Telescope/Multi Unit Spectroscopic Explorer (VLT/MUSE) to mosaic the central disk of galaxies with optical IFU observations within 
a 1$\farcm$0 × 1$\arcmin$0 field of view with  0$\farcs$2  pixels and a typical spectral resolution of $\sim$2.5{\AA}  over the wavelength range covering
4800-9300{\AA} \citep{Kreckel2019}. The angular resolution  between galaxies varies from 0$\farcs$5 to 1$\farcs$0. 

An integral field spectroscopic survey of a sample of 30 nearby spiral galaxies with the Mitchell Spectrograph (formerly called VIRUS-P) IFU on the 2.7 metre telescope
at McDonald Observatory, was carried out within the framework of the VIRUS-P Exploration of Nearby Galaxies (VENGA) survey \citep{Blanc2013,Kaplan2016}. The Mitchell Spectrograph has large
field of view (1$\farcm$7 × 1$\farcm$7). Each fibre is 4$\farcs$2 in diameter, the spatial resolution is 5.6 arcsec (full width half-maximum, FWHM).  Each galaxy was observed with both
a blue (3600–5800 \AA) and red (4600–6800 \AA) setup to obtain a wide wavelength coverage. The spectral resolution is R $\sim$1000 at 5000 \AA, which corresponds to $\sim$120 km/s.
Using the same instrument, the IFU spectroscopy of THINGS galaxies \citep{Walter2008}  is under observational campaigns as part of the  Metal-THINGS survey \citep{LaraLopez2021,LaraLopez2022}.
Some galaxies are observed with both a blue and red setups, however the red spectra is observed for all galaxies.

It is widely accepted that the direct T$_{e}$ method produces reliable estimations of the abundances in H\,{\sc ii} region. The auroral lines necessary for the application of the direct
T$_{e}$ method for the abundance determinations are measured in the long slit spectra of H\,{\sc ii} regions.  The amount of spectra of H\,{\sc ii} regions with detected auroral lines is
steadily growing. Nevertheless the use of the different variants of the strong line method (calibrations) is the dominant way for the estimations of the abundances in the interstellar
gas in galaxies. The  H\,{\sc ii} regions  with abundances determined through the direct method and the strong emission line fluxes measured in their spectra are used as the calibrating
data points in the construction of the empirical calibrations \citep[e.g.][]{Pilyugin2000,Pilyugin2001b,Pettini2004,Pilyugin2005,Pilyugin2011,Marino2013,Pilyugin2016,Curti2017}. Since
those empirical calibrations are based on the slit spectra then the applicability of that calibrations for the abundance determinations using the IFU spectra can be questioned.

Indeed, a spatial resolution and an angular size of the native spatial sample (fibre or spaxel) can significantly exceed the angular size of H\,{\sc ii} regions.
If this is the case then the diffuse ionized gas (DIG) outside  H\,{\sc ii} regions can make a contribution to the fibre (spaxel) spectrum as well to the extracted
spectrum of the  H\,{\sc ii} region.  However, \citet{Mannucci2021}  found that the difference between the spectra  local H\,{\sc ii} regions and more distant galaxies
is not due to contamination from the DIG, but by the smaller angular size of the slit with respect to the projected size of the  H\,{\sc ii}  regions,  and hence the DIG has a secondary effect.
Indeed, H\,{\sc ii} regions are stratified in the sense that higher ionization species dominate the inner regions of nebulae while lower ionization
species are more abundant in the outer parts (see Figure 1 in \citet{Mannucci2021} or Figure 3 in \citet{PerezMontero2014}). Therefore, the spectra from small apertures
show differences from the spectra when larger apertures are used, even if still inside the  H\,{\sc ii} regions. 

The goal of the current work is to examine the compatibility or disagreement between the calibration-based abundances obtained using the slit spectra of H\,{\sc ii}  regions and
the IFU spectra of spatial samplings (fibres) or extracted IFU spectra of H\,{\sc ii} regions. The three-dimensional R calibration (O/H)$_{R}$ = $f$(R$_2$,N$_2$,R$_3$/R$_2$)
and S calibration (O/H)$_{S}$ = $f$(N$_2$,S$_2$,R$_3$/S$_2$) from \citet{Pilyugin2016} are considered.  There is no one-to-one correspondence between positions and apertures of the
slit and IFU sampling. We determine the median value for abundances  in bins of 0.1 in R/R$_{25}$ and the scatter of each bin for both the slit and the IFU spectra-based abundances.
The use of median value in bins (the abundances at a given galactocentric distances) provides a possibility to compare the slit and IFU spectra-based abundances.    

Throughout the paper, we will use the following standard notations for the line intensities: \\ 
R$_2$  = $I_{\rm [O\,II] \lambda 3727+ \lambda 3729} /I_{{\rm H}\beta }$,  \\
N$_2$  = $I_{\rm [N\,II] \lambda 6548+ \lambda 6584} /I_{{\rm H}\beta }$,  \\
S$_2$  = $I_{\rm [S\,II] \lambda 6717+ \lambda 6731} /I_{{\rm H}\beta }$,  \\
R$_3$  = $I_{{\rm [O\,III]} \lambda 4959+ \lambda 5007} /I_{{\rm H}\beta }$.  \\
We also use the standard notation for electron temperature $t_{e}$ = 10$^{-4}T_{e}$. The galactocentric distance of  H\,{\sc ii} region or spatial sampling is given as a fractional
radius $R_{g}$ normalized to the optical radius $R_{25}$ of a galaxy, $R_{g}$ = $R$/$R_{25}$.

\section{Oxygen abundances in galaxies based on spectral measurements of different types}

\subsection{NGC~5457: comparison between (O/H)$_{T_{e}}$, (O/H)$_{R}$, and (O/H)$_{S}$ abundances based on the slit spectra}

\begin{figure}
\resizebox{1.00\hsize}{!}{\includegraphics[angle=000]{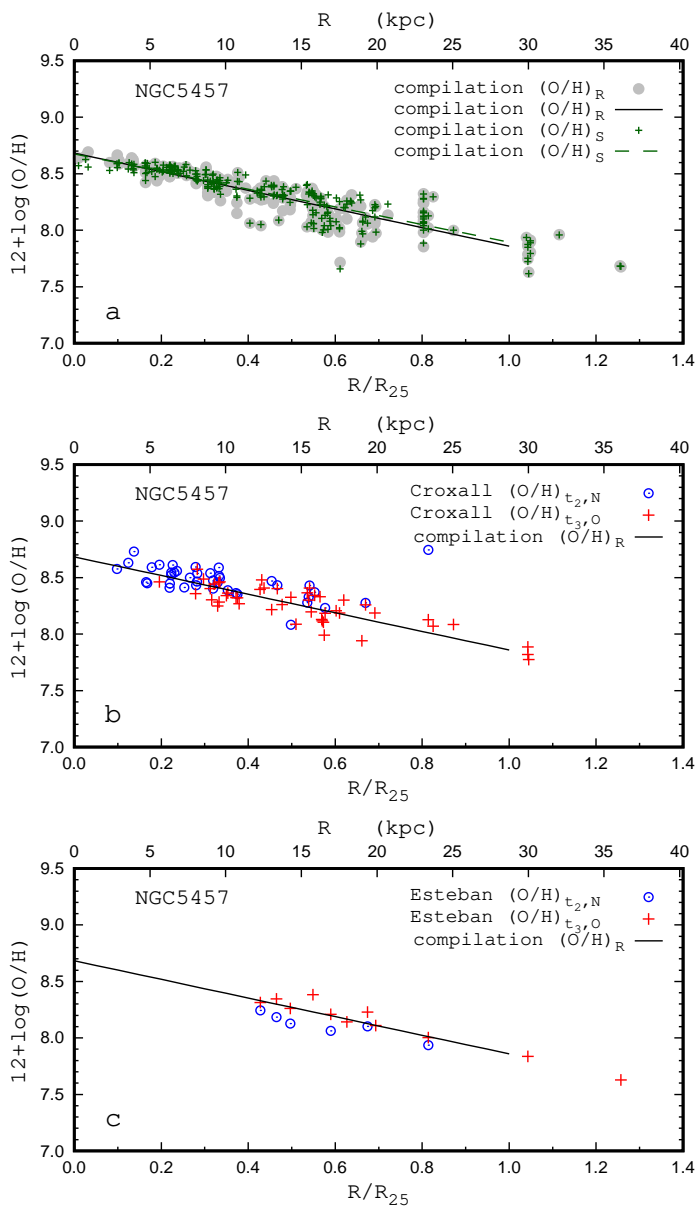}}
\caption{
  Radial oxygen abundance distribution in NGC~5457.
  {\sl Panel {\bf a}:} calibration-based oxygen abundances for the compilation of the slit spectra of H\,{\sc ii} regions.
  The grey circles denote the (O/H)$_{R}$ abundances for individual measurements, the black solid line is the best linear  fit to those data.
  The green plus signs are  the  (O/H)$_{S}$ abundances for individual measurements, the dashed line is the best linear  fit to those data. 
  {\sl Panel {\bf b}:} T$_{e}$-based oxygen abundances with measured electron temperatures $t_{2,N}$ (circles) and $t_{3,O}$ (plus signs)
  in H\,{\sc ii} regions from  \citet{Croxall2016}. The solid line is the (O/H)$_{R}$ -- $R$ relation from panel a.
  {\sl Panel {\bf c}:} the same as panel b but for the sample of H\,{\sc ii} regions from  \citet{Esteban2020}.
}
\label{figure:ngc5457-r-oh}
\end{figure}

The giant nearby galaxy NGC~5457 (M~101, the Pinwheel) is a prototype of the Sc spiral galaxies (morphological type code T = 5.9$\pm$0.3). 
 NGC~5457 is a face-on galaxy, its inclination angle is $i$ = 18$\degr$ and the position angle of the major axis PA = 37$\degr$ \citep{Kamphuis1993}.  
The optical radius of  NGC~5457 is R$_{25}$ = 14.42 arcmin or 865.2 arcsec \citep{RC3}. 

There are 79 independent distance measurements for NGC~5457 after the year 2000 including those using Cepheids and tip of the red giant branch (TRGB) \citep{Lomeli2022}. 
The obtained distances are within the range  $\sim$6 to $\sim$9 Mpc. We adopt here the distance to NGC~5457 d = 6.85. Each characteristic
of  NGC~5457 is rescaled if necessary to the distance adopted here. The optical radius of  NGC~5457 is R$_{25}$ = 28.73 kpc with the adopted distance. 

The oxygen abundance distribution across the disk of  NGC~5457 is considered in a number of investigations
\citep[][and references therein]{Kennicutt1996,Kennicutt2003,Pilyugin2001a,Croxall2016,Esteban2020}. Since the auroral lines were measured in the spectra of a number of
its H\,{\sc ii} regions and, consequently, the abundances in those H\,{\sc ii} regions can be derived through the direct $T_{e}$ method, then  NGC~5457 has been used
to test the validity of the calibrations, i.e., in some sense  NGC~5457 can been considered as a ``Rosetta stone'' \citep{Pilyugin2016}. Recent measurements of
H\,{\sc ii} regions spectra (including the auroral lines) in  NGC~5457 are published by \citet{Croxall2016} and  \citet{Esteban2020}. This provides an additional
possibility to compare the abundances produced by the empirical calibrations with the direct abundances. It should be noted that the measurements from \citet{Croxall2016} and
\citet{Esteban2020} were not used in constructing of the R and S calibrations, however, the spectra of H\,{\sc ii} regions of NGC~5457 with detectable auroral lines from other
publications were used.

The upper panel of Fig.~\ref{figure:ngc5457-r-oh} shows the oxygen abundances estimated through the R and S calibration from \citet{Pilyugin2016} as a function of radius for
the compilation of the slit spectra of H\,{\sc ii} regions in  NGC~5457 (measurements from \citet{Croxall2016} and  \citet{Esteban2020} are added to the compilation in
\citet{Pilyugin2016}). The grey points denote the (O/H)$_{R}$ abundances of individual H\,{\sc ii} regions, the black solid line is the  best linear fit  to those data. The green
plus signs are  the (O/H)$_{S}$ abundances of individual H\,{\sc ii} regions, the dashed line is the best fit to those data.

It is believed that the $T_{e}$ method, based on the measurements of temperature-sensitive line ratios, should give accurate oxygen abundances. In practice,
however, $T_{e}$-based oxygen abundances in the same H\,{\sc ii} region derived in different works can differ because of two reasons. First, there may be
errors in the line intensity measurements. Second,  the $T_{e}$-based oxygen abundances depend on the realization of the $T_{e}$ method, i.e. the derived
abundances depend on the relations used to convert the values of the line fluxes to the electron temperatures and to the ion abundances. The determined 
abundances depend also on the adopted relationship between electron temperature in low-ionization zones and electron temperatures in high-ionization parts of
nebula. The differences between the oxygen abundances at a given H\,{\sc ii} region produced by different realizations of the $T_{e}$ method can be appreciable 
\citep{Yates2020,Cameron2021}. For example, \citet{Esteban2017}  found an oxygen abundance of 12 + log(O/H)$_{T_{e}}$ = 8.14$\pm$0.05 in the Galactic H\,{\sc ii}
region Sh 2-83, while \citet{Arellano2020} found 12 + log(O/H)$_{T_{e}}$ = 8.28$\pm$0.08 in the same H\,{\sc ii} region using the same spectroscopic measurements. 
\citet{Berg2015}  detected the auroral lines in 45  H\,{\sc ii} regions in the nearby galaxy NGC~628. They determine the $T_{e}$-based abundances in those
H\,{\sc ii} regions and estimate the radial abundance gradient. They found the central (intersect) oxygen abundance 12 + log(O/H)$_{0}$ = 8.83$\pm$0.07
in the NGC~628. However, in their recent paper, \citet{Berg2020} recalculate ionic and total $T_{e}$-based abundances in those   H\,{\sc ii} regions
and determine the central (intersect) oxygen abundance 12 + log(O/H)$_{0}$ =  8.71$\pm$0.06 for the same galaxy.

The calibration-based abundances in H\,{\sc ii} regions of NGC~5457 are determined above through the R and S calibrations from \citet{Pilyugin2016}. The $T_{e}$-based
oxygen abundances in H\,{\sc ii} regions used as the calibrating data points in the construction of those calibrations are derived using the $T_{e}$-method equations reported
in \citet{Pilyugin2012}. In order for the calibration-based and the $T_{e}$-based abundances correspond to a unique abundance scale, the  $T_{e}$-based oxygen abundances for the
H\,{\sc ii} regions in  NGC~5457 are determined here using the $T_{e}$-method equations from \citet{Pilyugin2012}. 
It should be emphasized that we do not pretend that the T$_{e}$-based abundances in NGC~5457 recomputed here are more accurate than the original abundances in the cited papers. 
We only take care that all the abundances used here correspond to the same metallicity scale.  
If the measurements of two  auroral lines  ([O\,{\sc iii}]$\lambda$4363 and [N\,{\sc ii}]$\lambda$5755 are available for the H\,{\sc ii}
region then two values of the electron temperature ($t_{3,O}$ and $t_{2,N}$) can be derived and, consequently, two independent values of the oxygen abundance
((O/H)$_{t_{3},O}$ and (O/H)$_{t_{2},N}$) can be estimated. If the electron temperature in the low-ionized part of the nebula $t_{2}$ = $t_{2,N}$ is measured then the electron
temperature in the high-ionozation part of the nebula $t_{3}$ = $t_{3,O}$ is obtained using the relationships between electron temperatures in the nebula, and  conversely. 

Panel b of Fig.~\ref{figure:ngc5457-r-oh} shows the $T_{e}$-based oxygen abundances of NGC~5457 H\,{\sc ii} regions from  \citet{Croxall2016}. The oxygen abundances obtained
using the measured electron temperatures $t_{2,N}$ are shown by circles and those for the measured electron temperatures $t_{3,O}$ by plus signs. The (O/H)$_{R}$ -- R$_{g}$ relation
from panel a of Fig.~\ref{figure:ngc5457-r-oh} is shown by the solid line. Panel c of Fig.~\ref{figure:ngc5457-r-oh} shows the oxygen abundances  of NGC~5457 H\,{\sc ii} regions
from  \citet{Esteban2020}. 

Inspection of the Fig.~\ref{figure:ngc5457-r-oh} suggests that the radial abundance gradients traced by the $T_{e}$-based,  R and S calibration-based abundances determined from
the slit spectra are in a satisfactory agreement to each other. Therefore the slit spectra-based abundances determined through the R and S calibrations can be used 
as the reference abundances.

\subsection{NGC~628: comparison between abundances based on slit and IFU spectra}

\subsubsection{Slit spectra-based abundances in NGC~628}

\begin{figure}
\resizebox{1.00\hsize}{!}{\includegraphics[angle=000]{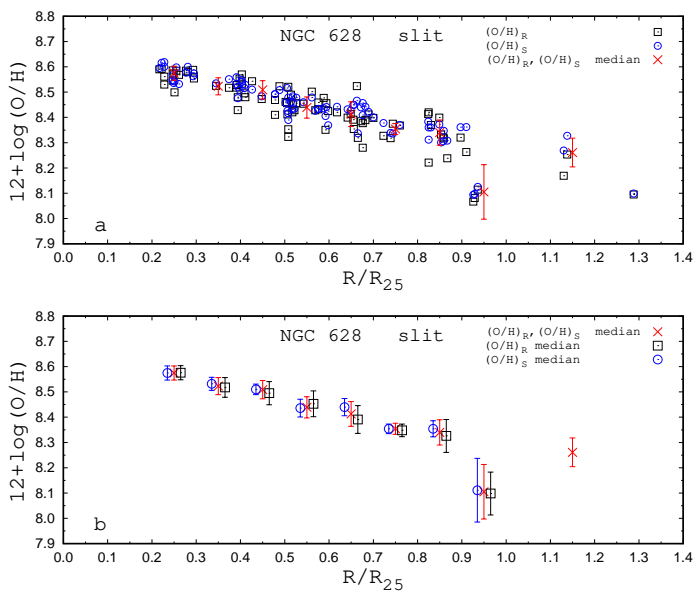}}
\caption{
  Radial distribution of the slit spectra-based oxygen abundances in NGC~628.
  {\sl Panel {\bf a}:} squares denote the abundances estimated through the R calibration for individual measurements, circles are the S calibration-based abundances. 
  The crosses mark the median value for (O/H)$_{R}$ and (O/H)$_{S}$ together in bins of 0.1 in R/R$_{25}$ and the scatter of each bin.
  {\sl Panel {\bf b}:}  the median value for the (O/H)$_{S}$ (circles), for the (O/H)$_{R}$ (squares), and for the (O/H)$_{R}$ and the (O/H)$_{S}$ together (crosses).
  The median values for different abundances are obtained for the same bin, but the positions of symbols (circles and squares) are slightly shifted along the X axis for clarity.
}
\label{figure:ngc0628-r-oh-slit}
\end{figure}

\begin{figure}
\resizebox{1.00\hsize}{!}{\includegraphics[angle=000]{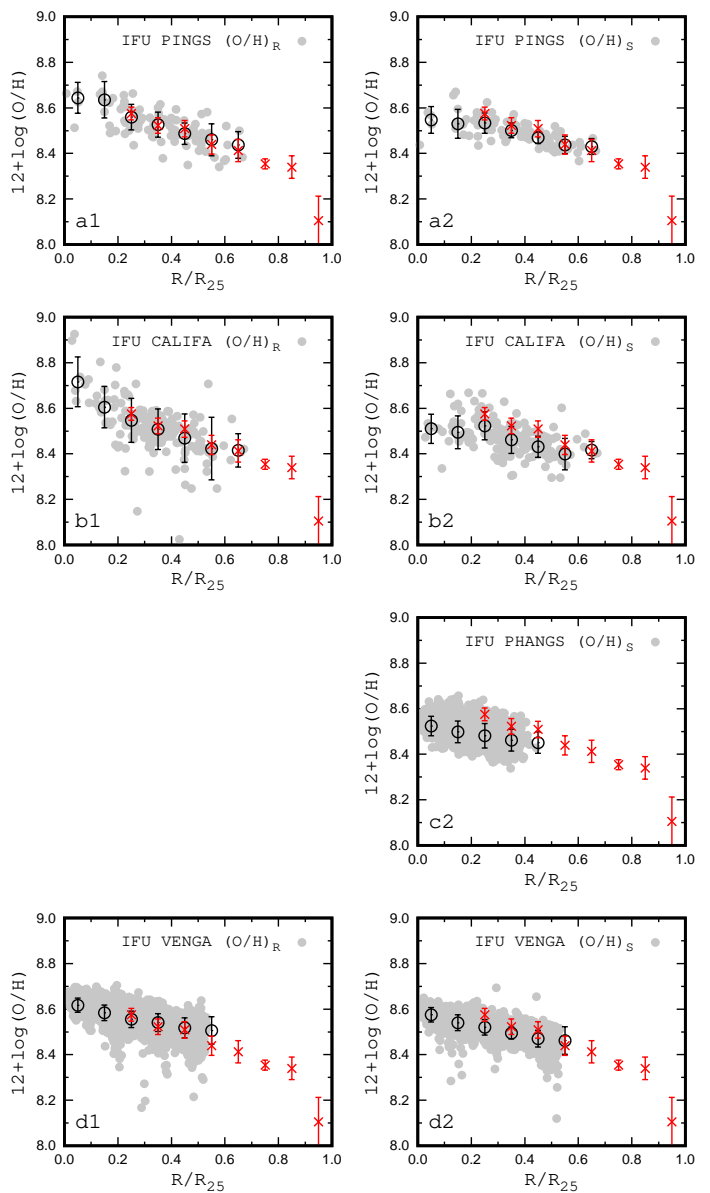}}
\caption{
  Radial distributions of the oxygen abundances in NGC~628.
  {\sl Panels of columns 1-2:} radial distributions of abundances estimated through the R calibration (column 1)  and S calibration (column 2). 
  {\sl Panels of rows a-d:} abundances based on the  IFU measurements from PINGS (row a), from CALIFA survey (row b), from PHANGS survey (row c), and from VENGA survey (row d).
  The grey points denote the abundances for the individual H\,{\sc ii} regions (panels of rows a, b, c) or individual spatial samplings (panels of row d),
  the dark circles mark the median value in  bins   of 0.1 in R/R$_{25}$ and the scatter of each bin. The red crosses in each panel designate the median values for the slit
  spectra-based abundances ((O/H)$_{R}$ and (O/H)$_{S}$ together) and comes from  Fig.~\ref{figure:ngc0628-r-oh-slit}.   
}
\label{figure:ngc0628-r-oh-22}
\end{figure}

\begin{figure}
\resizebox{1.00\hsize}{!}{\includegraphics[angle=000]{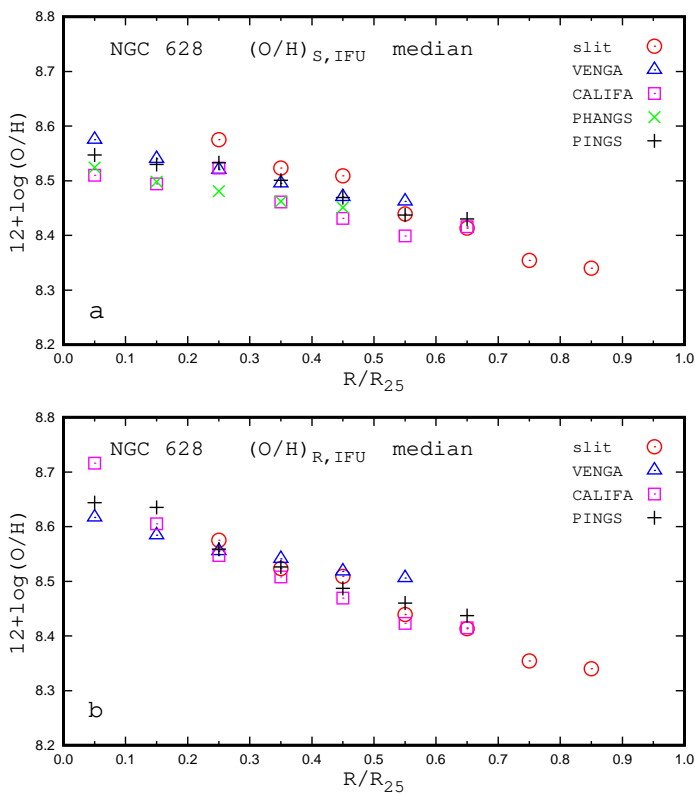}}
\caption{
  The median values of oxygen abundances in bins of NGC~628.
  {\sl Panel {\bf a}:} median values of (O/H)$_{S}$ abundances  based on the IFU measurements from PINGS (plus signs), from CALIFA survey (squares),
  from PHANGS survey (crosses), and from VENGA survey (triangles).
  The median values of the abundances  based on the slit measurements (circles) are obtained for the (O/H)$_{S}$ and (O/H)$_{R}$ abundances  together.
  {\sl Panel {\bf b}:} the same as panel a but for the (O/H)$_{R}$ abundances.
}
\label{figure:ngc0628-r-oh-median}
\end{figure}

%
\begin{figure*}
\resizebox{1.00\hsize}{!}{\includegraphics[angle=000]{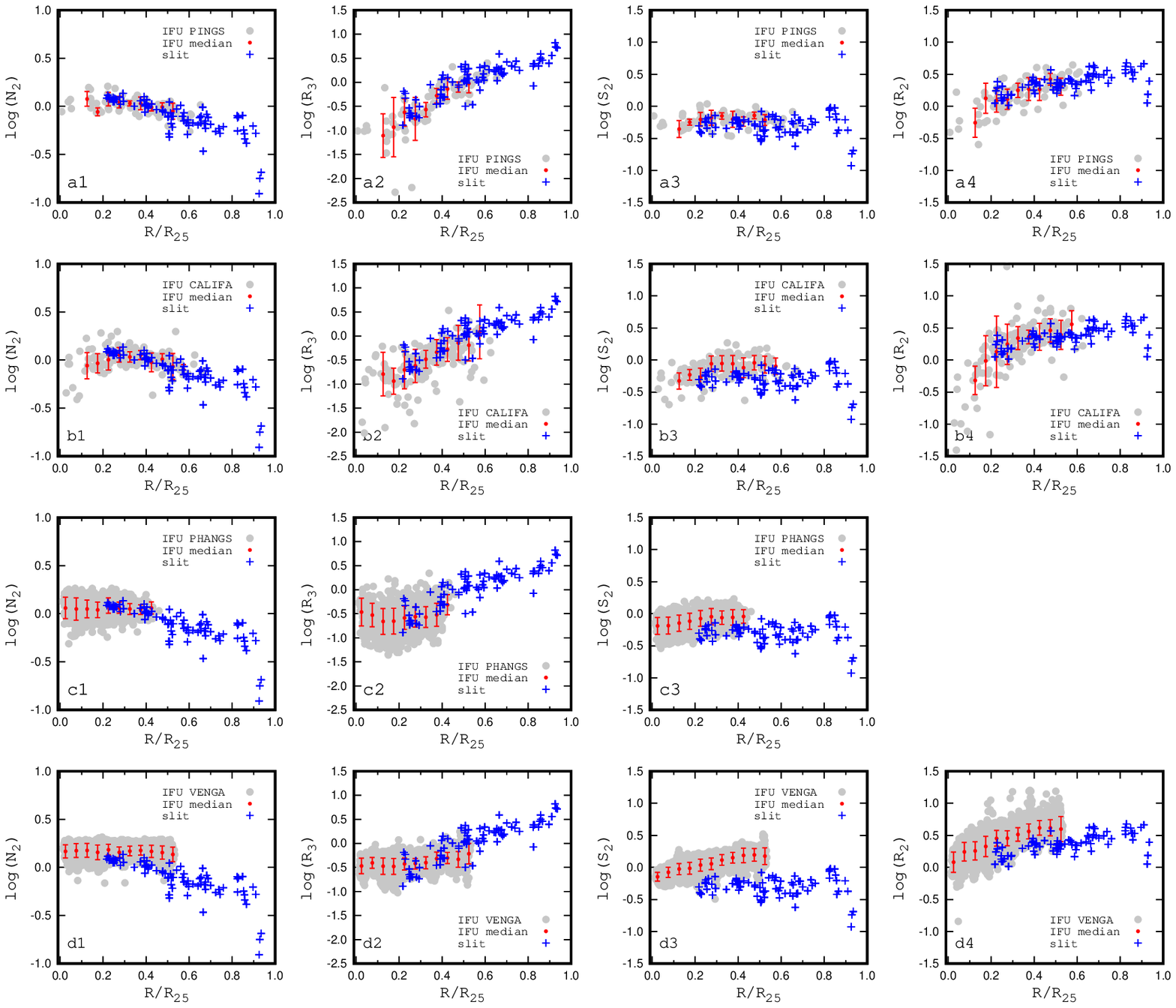}}
\caption{
  Radial distributions of the emission line fluxes in NGC~628.
  {\sl Panels of columns 1-4:} radial distributions of intensities of N$_{2}$ line (column 1), R$_{3}$ line (column 2), S$_{2}$ line (column 3), and R$_{2}$ line (column 4). 
  {\sl Panels of rows a-d:} IFU measurements from the surveys PINGS (row a),  CALIFA  (row b),  PHANGS  (row c), and  VENGA  (row d).
  The grey points denote the individual H\,{\sc ii} regions (panels of rows a, b, c) or individual spatial samplings (panels of row d), the red points with bars mark the median value in  bins
  of 0.05 in R/R$_{25}$ and the scatter of each bin. The blue plus signs are individual slit measurements. 
  Since only the red spectra were measured within the PHANGS programme then the R$_{2}$ line measurements are not available. 
}
\label{figure:ngc0628-r-line}
\end{figure*}

The nearby galaxy NGC~628 (M~74, the Phantom Galaxy) is an isolated Sc spiral galaxy (morphological type code T = 5.2$\pm$0.5).  NGC~628 is a face-on galaxy,
its inclination angle is $i$ = 6$\degr$ and the position angle of the major axis PA = 25$\degr$ \citep{Kamphuis1992}. The optical radius of  this galaxy is R$_{25}$ = 5.24
arcmin or 314.1 arcsec \citep{RC3}.  The  H\,{\sc i} disk extends out to more than three times the optical radius \citep{Kamphuis1992}. 
There are recent distance estimations for  NGC~628 through the tip of the red giant branch (TRGB) method based on the Hubble Space Telescope measurements. 
\citet{Jang2014} find the distance to NGC~628 to be 10.19 $\pm$ 0.14 (random) $\pm$ 0.56 (systematic) Mpc. \citet{McQuinn2017} measure the distance to M74 to be
9.77 $\pm$ 0.17 (statistical uncertainty) $\pm$ 0.32 (systematic uncertainty) Mpc. \citet{Sabbi2018} determine the distances for the central pointing (d = 8.6 $\pm$ 0.9 Mpc)
and for the outer field (d = 8.8 $\pm$ 0.7 Mpc). We adopt here the distance to the NGC~628 used in our previous paper (d = 9.91 Mpc \citep{Pilyugin2014}) which is close
to the values obtained by \citet{Jang2014} and \citet{McQuinn2017}.  The optical radius of  NGC~628 is R$_{25}$ = 15.09 kpc with adopted distance. 

The slit spectra of the H\,{\sc ii} regions in the NGC~628 were measured in a number of early works \citep{McCall1985,Ferguson1998,vanZee1998,Bresolin1999}, and the slit spectroscopy of
H\,{\sc ii} regions in NGC~628 were carried out within the framework of the CHemical Abundances Of Spirals (CHAOS) project \citep{Berg2015}.

Squares in panel a of Fig.~\ref{figure:ngc0628-r-oh-slit} show the oxygen abundances estimated through the R calibration in individual H\,{\sc ii} regions and circles are the S
calibration-based abundances. The median values of the (O/H)$_{R}$ and the (O/H)$_{S}$ together in bins of 0.1 in R/R$_{25}$ are denoted by crosses, the bar marks the scatter in each bin.
Panel b of Fig.~\ref{figure:ngc0628-r-oh-slit} show the comparison between median values for the (O/H)$_{S}$ (circles), for the (O/H)$_{R}$ (squares), and for the (O/H)$_{R}$ and the
(O/H)$_{S}$ together (crosses). The median values for different abundances are obtained for the same bin, but the positions of symbols (circles and squares) are slightly shifted
along the X axis for clarity. Inspection of panel b of Fig.~\ref{figure:ngc0628-r-oh-slit} shows that the differences between median value for (O/H)$_{S}$ (or (O/H)$_{R}$) abundances
and the median value for the (O/H)$_{S}$ and the (O/H)$_{R}$ abundances together is less than the uncertainties of median values. This justifies the use of the median value for the
(O/H)$_{S}$ and the (O/H$_{R})$ abundances together instead of the median values for the (O/H)$_{S}$ or the (O/H)$_{R}$ abundances separately to spesify the abundance in bin in the case
of the slit spectra-based abundances.  The advantages of such approach are the following. First. In this case, the IFU spectra-based (O/H)$_{S}$ and (O/H)$_{R}$ abundances are compared
with the same reference (slit spectra-based) abundances. Second. The slit spectra measurements for a galaxy are usually a few in number (see below). Therefore, the use of  (O/H)$_{S}$
and (O/H)$_{R}$ abundances together provides a possibility to estimate the median values in larger number of bins than for the (O/H)$_{S}$ or the (O/H)$_{R}$ abundances only. 

Thus, in the case of the slit spectra-based abundances, we will estimate and use the median value in the bin for the (O/H)$_{S}$ and the (O/H$_{R})$ abundances together and will
notate those median values as (O/H)$_{SLIT}$.

\subsubsection{IFU spectra-based abundances in NGC~628}

\citet{Rosales2011} identify  H\,{\sc ii} regions and extract their spectra from the IFU measurements. They generate the  H\,{\sc ii} regions catalog for NGC~628 (PINGS catalog). The oxygen
abundances in the H\,{\sc ii} regions catalog from \citet{Rosales2011} are obtained via the R calibration are denoted by grey circles in panel a1 of Fig.~\ref{figure:ngc0628-r-oh-22}.
The dark circles mark the median value in  bins   of 0.1 in R/R$_{25}$ and the scatter of each bin. The red crosses designate the median values in bins for the slit
spectra-based abundances ((O/H)$_{R}$ and (O/H)$_{S}$ together) and comes from  Fig.~\ref{figure:ngc0628-r-oh-slit}.   
Panel a2 of Fig.~\ref{figure:ngc0628-r-oh-22} shows the oxygen abundances in the H\,{\sc ii} regions from  \citet{Rosales2011}
estimated through the S calibration (grey points), and the dark circles mark the median value in  bins and the scatter of each bin.

\citet{Espinosa2020} have also constructed a catalog of H\,{\sc ii} regions in the NGC~628 using IFU observations (CALIFA catalog). The oxygen abundance in H\,{\sc ii} regions of
the CALIFA catalog obtained through the R calibration are denoted by grey circles in panel b1 of Fig.~\ref{figure:ngc0628-r-oh-22}, The dark circles are the median values in  bins,
and the bar shows the scatter of each bin. The  grey circles in panel b2 of Fig.~\ref{figure:ngc0628-r-oh-22} are the oxygen abundances in those H\,{\sc ii} regions estimated through
the S calibration and the dark circles are the median values in  bins.

The IFU spectroscopy of the central part of  NGC~628 were carried out within the framework of the PHANGS programme using the Very Large Telescope \citep{Kreckel2019}. The catalog 
of the H II regions in  NGC~628 was created on the base of those measurements. Unfortunately only the red spectra  over the wavelength range covering 4800-9300{\AA} were obtained.
Since the line [O\,{\sc ii}]$\lambda\lambda$3727,3729 is not measured then the R calibration cannot be applied to those  H II regions.  The grey points in panel c2 of
Fig.~\ref{figure:ngc0628-r-oh-22} denote  the oxygen abundances in individual H\,{\sc ii} regions estimated through the S calibration,  and the dark circles are the median values in  bins.

The IFU (fibre) spectroscopy of  NGC~628  was performed within the framework of the VENGA survey \citep{Blanc2013,Kaplan2016}.  The oxygen abundance in the individual fibres obtained through
the R calibration are denoted by grey circles in panel d1 of Fig.~\ref{figure:ngc0628-r-oh-22}. The grey circles in panel d2 of Fig.~\ref{figure:ngc0628-r-oh-22} are the oxygen abundances
in the fibres estimated through the S calibration.   The dark circles in both panels mark the median value in  bins and the scatter of each bin.

We compare the abundances in NGC~628 obtained from IFU spectra from different surveys to each other and to the slit spectra-based abundances in Fig.~\ref{figure:ngc0628-r-oh-median}.  
Panel a shows the median values of (O/H)$_{S}$ abundances  based on the IFU measurements from PINGS (plus signs), from CALIFA survey (squares), from PHANGS survey (crosses), and from
VENGA survey (triangles). The median values of the abundances  based on the slit measurements (circles) are obtained for the (O/H)$_{S}$ and (O/H)$_{R}$ abundances  together.
Panel b shows a similar comparison for the (O/H)$_{R}$ abundances.

\citet{Easeman2022} find that central dips in the metallicity profiles within galaxies can be observed using spatially resolved IFU data. It is not clear whether the dips are real
or they are an artefact introduced by the strong line diagnostics used to determine the metallicity. The radial changes of the (O/H)$_{S}$ abundances based on the PINGS and the CALIFA IFUs
measurements are not perfectly monotonic, panels a2 and b2 in  Fig.~\ref{figure:ngc0628-r-oh-22}. However, it is not clear whether the deviation from the monotonoc trend is caused by
the decrease in the central metallicity (at $R/R_{25} \la$  0.2) or by the enhancement of the metallicity in the only  bin, at $R/R_{25}$ = 0.2 to 0.3. If this bin is ignored then the
metallicity profile becomes monotonic. Moreover the radial changes of the (O/H)$_{S}$ abundances based on the PHANGS and the VENGA IFUs measurements are monotonic, panels c2 and d2 in
Fig.~\ref{figure:ngc0628-r-oh-22} fig.3. Thus, the central dip in the metallicity profile within the NGC~628 is unlikely.

Inspection of the  panels of columns 1 and 2 in Fig.~\ref{figure:ngc0628-r-oh-22} and Fig.~\ref{figure:ngc0628-r-oh-median}  shows  the following. \\
-- The (O/H)$_{R}$ abundances based on the IFU spectra of  H\,{\sc ii} regions from PINGS  and CALIFA  and spectra of fibres from VENGA are in  agreement
to each other and to the slit spectra-based abundances. \\
-- The (O/H)$_{S}$ abundances based on the IFU spectra of H\,{\sc ii} regions from PINGS, CALIFA, and PHANGS and on the IFU spectra of fibres from VENGA  are more or
less similar. The IFU spectra-based (O/H)$_{S}$ abundances are close to the slit spectra-based abundances at large galactocentric distances but there is difference
at smaller galactocentric distances (at high metallicities) in the sense that the IFU spectra-based (O/H)$_{S}$ abundances are lower than the slit spectra-based abundances.
As a result, the radial abundance gradient traced by the (O/H)$_{S}$ abundances based on the IFU  spectra  is flatter than that for abundances based on the slit spectra.
\citet{Belfiore2017}  find that the galaxy inclination generates a flattening on the radial abundance gradient, since flux from different galactocentric radii is summed up when
the galaxy is projected in the plane of the sky. The metallicity depletion at the galaxy centre depends on the PSF and galaxy inclination and can be as large as $\sim$0.02 dex.
Since NGC~628 is a face-on galaxy (with inclination $i$ = 6$\degr$) then this effect should be minor.   \\
Unfortunately, the slit spectra measurements are not available at the central region ($R/R_{25} \la$ 0.2) of NGC~628 while the IFU spectra are not available at larger radii
($R/R_{25} \ga$ 0.65). Therefore, the comparison is possible within the restricted interval of galactocentric distances only. 

Comparison between the emission line fluxes in the slit and IFU spectra in  NGC~628 can clarify the origin of the difference in abundances.
Fig.~\ref{figure:ngc0628-r-line} shows the  radial distributions of the emission line fluxes in the slit and the IFU spectra in  NGC~628. The grey points denote the individual IFU  spectra
of H\,{\sc ii} regions (panels of rows a, b, and c) or individual fibres (panels of row d), the red points with bars mark the median value in the bins of 0.05 in $R/R_{25}$ and  the scatter
of each bin. The blue plus signs are individual slit spectra. The differences between  the emission line fluxes in the slit and IFU spectra result in the differences between the slit
spectra-based and the IFU spectra-based abundances.

The difference in S$_{2}$ (and other lines) between different IFU measurements cannot be attributed to the difference in the spatial resolution only. If this would be the case then one
could expect that the S$_{2}$ for a given galaxy (e.g. NGC 628) should be similar in the CALIFA and PINGS IFUs since the spatial and spectral resolutions are the same. But there is an
appreciable difference between S$_{2}$  from those surveys (compare panels a3 an b3 in Fig.~\ref{figure:ngc0628-r-line}). Then one can assume that the reduction of the observations
(e.g. extraction of  H\,{\sc ii} region spectra) and flux calibration, can contribute to the uncertainty in the IFU spectra and, as consequence, can produce some differences in the S$_{2}$
measurements between different IFUs. However, the general behaviour of the IFU spectra-based abundances are more or less similar for different IFUs.  

It may appear that  (O/H)$_{S,IFU}$ -- (O/H)$_{SLIT}$ is larger than  (O/H)$_{R,IFU}$ -- (O/H)$_{SLIT}$ (Fig.~\ref{figure:ngc0628-r-oh-22} and Fig.~\ref{figure:ngc0628-r-oh-median})  because of the
larger differences in the S$_{2}$ flux measured between the IFU and the slit spectra. However, close examination of the available data reveals that differences in other lines fluxes must
also play a significant role.
For example, we find that the differences in the S$_{2}$ flux between PINGS and VENGA are larger than between PINGS and CALIFA (Fig.~\ref{figure:ngc0628-r-line}), but the differences in (O/H)$_{S}$
between PINGS and VENGA are smaller than that between PINGS and CALIFA  (Fig.~\ref{figure:ngc0628-r-oh-22} and Fig.~\ref{figure:ngc0628-r-oh-median}). 
Moreover, at $R$/$R_{25}$ = 0.45 to 0.50, the median values of the R$_{2}$ and S$_{2}$ fluxes in CALIFA and PINGS agree within $\sim$10\% and $\sim$25\%, respectively, and the
difference between the median values of (O/H)$_{R}$ and (O/H)$_{S}$ are $\sim$0.01 dex and $\sim$0.05 dex, respectively. Whereas, at $R$/$R_{25}$ = 0.15 to 0.20, the median values
of the R$_{2}$ and S$_{2}$ fluxes in CALIFA and PINGS agree within $\sim$25\% and $\sim$10\%, respectively, but the difference between the median values of (O/H)$_{R}$ and
(O/H)$_{S}$ are $\sim$0.02 dex and $\sim$0.03 dex, respectively. This suggests that the abundances obtained through the S calibration are affected by variations in the line fluxes
more strongly than abundances obtained through the R calibration.

From one side, since the spatial resolutions (and sizes of the native spatial samplings) of PHANGS, CALIFA, PINGS, and VENGA observations are appreciable different then one can expect that
the spectra of the extracted H\,{\sc ii} regions in PHANGS should be less contaminated by the DIG than the spectra of the extracted H\,{\sc ii} regions in PINGS and CALIFA and than the spectra
of spatial samplings in VENGA. From other side, the shifts of the (O/H)$_{S}$ abundances based on the IFU spectra of PHANGS, CALIFA, PINGS, and VENGA in comparison to the slit spectra-based
abundances are rather close to each other. Those two facts taken together are in line with the conclusion of  \citet{Mannucci2021} that the DIG has a secondary effect.

Thus, the investigation of the galaxy NGC~628 can be summarized as following. 
{\it i)} the oxygen abundance estimated from the IFU spectra of the extracted H\,{\sc ii} regions, or from the IFU spectra of fibres, depends weakly on the spatial resolution (and size of the
native spatial samplings) of the IFU measurements. 
{\it ii)} the R calibration  applied to the IFU spectra produce more   consistent abundance estimations between slit and IFU spectra than the S calibration, the (O/H)$_{S,IFU}$ abundances
are systematically underestimated in comparison with the slit spectra-based abundances at high metallicities. 
{\it iii)}  the satisfactory agreement between abundances estimated from the IFU spectra obtained with different spatial resolutions (and different sizes of the native spatial samplings)
confirm the recent finding of  \citet{Mannucci2021} that the DIG has a secondary effect.
To confirm or reject those conclusions, seven other nearby galaxies with available slit and IFU spectroscopy are examined (see Appendix A).

\section{Discussion}

Here we compare and discuss the abundances based on the slit spectra and the abundances determined from the IFU spectra using the R and the S calibrations. 
There are 38 bins of 0.1 in R/R$_{25}$ in eight galaxies considered where the median values of both (O/H)$_{R,IFU}$ and (O/H)$_{SLIT}$ abundances are determined and 57 bins where the median
values of both (O/H)$_{S,IFU}$ and (O/H)$_{SLIT}$ abundances are determined (Fig.~\ref{figure:ngc0628-r-oh-22},  Fig.~\ref{figure:ngc1058-r-oh} -- Fig.~\ref{figure:ngc6946-r-oh}).  

\begin{figure}
\resizebox{1.00\hsize}{!}{\includegraphics[angle=000]{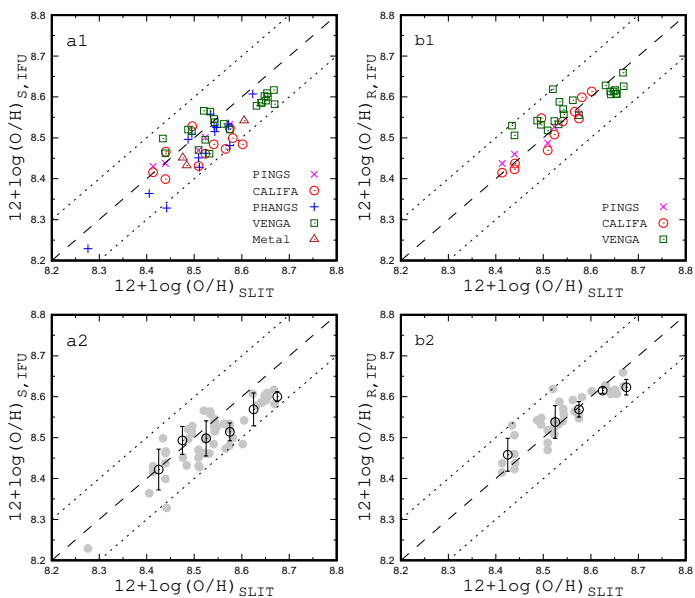}}
\caption{
  Comparison between the abundances based on the slit spectra  of H\,{\sc ii} regions, and the abundances estimated from the IFU spectra of H\,{\sc ii} regions (or spatial samplings, fibres)
  using the R and the S calibrations. 
  {\sl Panel {\bf a1}:} IFU spectra-based oxygen abundances (median values in bins) obtained through the S calibration vs. oxygen abundances estimated from the slit spectra
  for our sample of galaxies.
  Each point shows the median value in bin of 0.1 in R/R$_{25}$ in galaxy (Fig.~\ref{figure:ngc0628-r-oh-22},  Fig.~\ref{figure:ngc1058-r-oh} -- Fig.~\ref{figure:ngc6946-r-oh}).  
  The abundances based on the IFU spectra from different surveys are shown by different symbols. The solid line indicates the one-to-one relation;
  while the dashed lines show the $\pm$0.1 dex deviation.
  {\sl Panel {\bf b1}:} the same as panel a1 but for the IFU spectra-based abundances estimated through the R calibration.
  {\sl Panel {\bf a2}:} grey points are data from panel a1, the dark circles mark the mean value in  bins  of 0.05 in log(O/H) for those data points. 
  {\sl Panel {\bf b2}:} the same as panel a2 but for the IFU spectra-based abundances estimated through the R calibration.
  }
\label{figure:ohslit-ohifu}
\end{figure}

\begin{figure}
\resizebox{1.00\hsize}{!}{\includegraphics[angle=000]{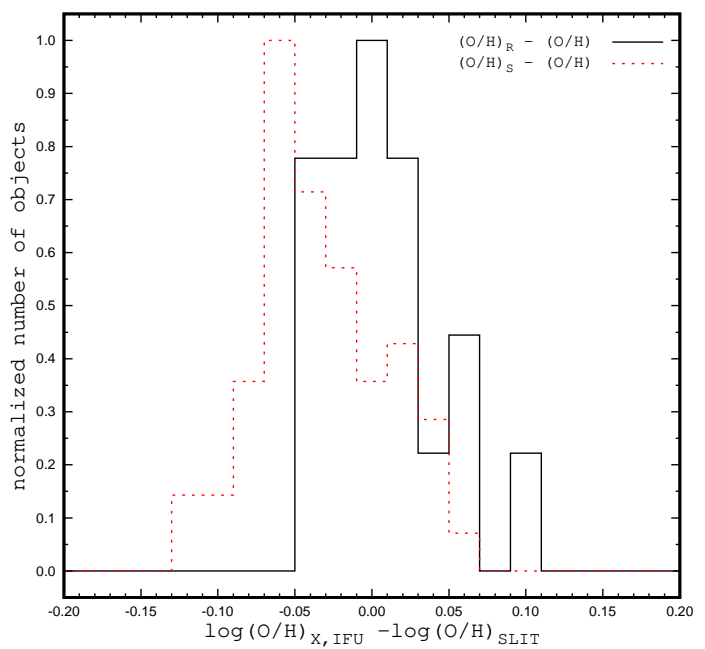}}
\caption{
  The normalized histograms of the differences between oxygen abundances (O/H)$_{R,IFU}$ -- (O/H)$_{SLIT}$ (solid line) and   (O/H)$_{S,IFU}$ -- (O/H)$_{SLIT}$ (dashed line).  
}
\label{figure:gist-ohslit-ohifu}
\end{figure}

In panel a1 of Fig.~\ref{figure:ohslit-ohifu} we plot the abundance determined from the IFU spectra through the S calibration in individual bins as a function of the abundance obtained
from the slit spectra   for all eight galaxies,  and in panel b1 of Fig.~\ref{figure:ohslit-ohifu} we plot the (O/H)$_{R,IFU}$ abundance against the (O/H)$_{SLIT}$ abundances.
The grey points in panel a2 of Fig.~\ref{figure:ohslit-ohifu} are the abundances in individual bins (the same data as in panel a1) and  the dark circles mark the mean value in  bins
of 0.05 in log(O/H)$_{SLIT}$ for those data points. Panel b2  of Fig.~\ref{figure:ohslit-ohifu} shows the diagram similar to that in panel a2 but for the (O/H)$_{R,IFU}$ abundances.

Fig.~\ref{figure:gist-ohslit-ohifu} shows the normalized histograms of the differences between oxygen abundances (O/H)$_{R,IFU}$ -- (O/H)$_{SLIT}$ for 38 data points (solid line) and between
(O/H)$_{S,IFU}$ -- (O/H)$_{SLIT}$ for 57 data points (dashed line). The step in the histograms is 0.02 dex.  
 
Inspection of panel a1 of Fig.~\ref{figure:ohslit-ohifu} and Fig.~\ref{figure:gist-ohslit-ohifu} shows that the differences between (O/H)$_{SLIT}$ and (O/H)$_{S,IFU}$ abundances are usually
less than around 0.1 dex, the mean value of the scatter in abundance differences (O/H)$_{S,IFU}$ -- (O/H)$_{SLIT}$ is 0.052 dex for the 57 data points. However, the differences are not perfectly
random, the maximum of the distribution is appreciable shifted from zero, Fig.~\ref{figure:gist-ohslit-ohifu}. Examination of  panels a1 and a2 of
Fig.~\ref{figure:ohslit-ohifu} shows that the (O/H)$_{S,IFU}$ abundances are systematically lower than the (O/H)$_{SLIT}$ abundances at high metallicities (12 + log(O/H) $\ga$ 8.55), the mean
shift of the (O/H)$_{S,IFU}$ abundances relative to the (O/H)$_{SLIT}$ abundances is --0.059 dex for 21 data points, that is the shift of  (O/H)$_{S,IFU}$ abundances  relative to the (O/H)$_{SLIT}$
abundances at high metallicities exceeds the mean value of the scatter in the abundance differences.

Inspection of panel b1 of Fig.~\ref{figure:ohslit-ohifu} and Fig.~\ref{figure:gist-ohslit-ohifu} shows that the differences between (O/H)$_{SLIT}$ and (O/H)$_{R,IFU}$ abundances are 
also within  0.1 dex,   the mean value of the scatter in abundance differences (O/H)$_{R,IFU}$ -- (O/H)$_{SLIT}$ is 0.037 dex for the 38 data points. One can observe that the correlation
between (O/H)$_{R,IFU}$ and (O/H)$_{SLIT}$ abundances is more tight than the correlation between (O/H)$_{S,IFU}$ and (O/H)$_{SLIT}$ abundances. The maximum of the abundance differences distribution
is close to zero,  Fig.~\ref{figure:gist-ohslit-ohifu}. Nevertheless, close examination of  panels b1 and b2 of Fig.~\ref{figure:ohslit-ohifu} shows that some correlation between the abundance
difference  (O/H)$_{R,IFU}$ -- (O/H)$_{SLIT}$  and metallicity can exists. However, the mean shift of the  (O/H)$_{R,IFU}$ abundances
relative (O/H)$_{SLIT}$ abundances exceeds the scatter in abundance differences for one metallicity interval (12 + log(O/H) from 8.65 to 8.70) only. Therefore it is not clear whether this
weak correlation is meaningful.

Inspection of panel a1 of Fig.~\ref{figure:ohslit-ohifu} shows that there is an agreement between the (O/H)$_{S,IFU}$ estimated from the IFU measurements obtained with different spatial
resolutions, and panel b1 shows such agreement for the (O/H)$_{R,IFU}$ abundances. This suggests that the IFU spectra-based abundances depend weakly (if any) on the spatial resolution
of the IFU measurements. Indeed, there is an agreement between the (O/H)$_{S,IFU}$ abundances estimated in NGC~4254 from the VENGA measurements (panel b in Fig.~\ref{figure:ngc4254-r-oh})
and from the PHANGS measurements (panel d in Fig.~\ref{figure:ngc4254-r-oh}) although the spatial resolution of the VENGA measurements is significantly lower than that of the PHANGS measurements.
Moreover, the agreement between (O/H)$_{S,IFU}$ abundances estimated in NGC~628 from the PINGS and the VENGA measurements obtained with different spatial resolutions is better than the   
agreement between (O/H)$_{S,IFU}$ abundances estimated from the PINGS and the CALIFA measurements obtained with the same spatial resolution, Fig.~\ref{figure:ngc0628-r-oh-median}. 

Thus, the analysis of a sample of nearby galaxies confirms the conclusions reached above when considering the galaxy NGC~628.
It should be emphasized that the abundances analised here are distributed over the limited interval of metallicity, from 12 + log(O/H) $\sim$8.4 to $\sim$8.7 only (with one exception). 
In order to strengthen (or reject) our conclusions and derive the relationships between the IFU and slit spectra-based abundances, one needs the abundances distributed over the larger
diapason of metallicity.

\section{Conclusions}

The calibration-based abundances are considered in eight nearby galaxies where both, the slit spectra of H\,{\sc ii} regions and the spectra of H\,{\sc ii} regions extracted
from the IFU measurements (PINGS, CALIFA, and PHANGS surveys) or the IFU spectra of spatial samplings (VENGA and Metal-THINGS surveys) are available.
The pre-existing empirical R and S calibrations for abundance determinations are constructed using a sample of H\,{\sc ii} regions with high quality slit spectra \citep{Pilyugin2016}.  
In this study, we test the applicability of those calibrations to the IFU spectra. 
We estimate the R and S calibration-based abundances using both the IFU and the slit spectroscopy for eight nearby galaxies. The median values of the IFU
spectra-based abundances (O/H)$_{R,IFU}$ and (O/H)$_{S,IFU}$ in bins of 0.1 in fractional radius $R_{g}$ (normalized to the optical radius $R_{25}$) of a galaxy are determined and compared to
the median values of the slit spectra-based abundances (O/H)$_{SLIT}$ determined using both the (O/H)$_{R}$ and (O/H)$_{S}$ abundances together.
 
We find that the differences between (O/H)$_{SLIT}$ and (O/H)$_{S,IFU}$ abundances are usually less than around 0.1 dex, the mean value of the scatter in abundance differences
(O/H)$_{S,IFU}$ -- (O/H)$_{SLIT}$ is 0.052 dex for the 57 data points. However, the (O/H)$_{S,IFU}$ abundances are systematically lower than the (O/H)$_{SLIT}$ abundances at high
metallicities (12 + log(O/H) $\ga$ 8.55), the mean shift of the (O/H)$_{S,IFU}$ abundances relative to the (O/H)$_{SLIT}$ abundances is around --0.06 dex for 21 data points. 
The difference between the intensities of each line (used in the abundance determinations) in the slit and the IFU spectra contributes to the difference between the abundances
based on the slit and the IFU spectra. Our data indicates that the abundances estimated through the S calibration are more sensitive to the variations in the line fluxes than the
abundances obtained using the R calibration.

The correlation between (O/H)$_{R,IFU}$ and (O/H)$_{SLIT}$ abundances is more tight than the correlation between (O/H)$_{S,IFU}$ and (O/H)$_{SLIT}$ abundances.   
the mean value of the scatter in abundance differences (O/H)$_{R,IFU}$ -- (O/H)$_{SLIT}$ is 0.037 dex for the 38 data points. There is hint that the weak correlation between the abundance
difference  (O/H)$_{R,IFU}$ -- (O/H)$_{SLIT}$  and metallicity can exists. However, the mean shift of the  (O/H)$_{R,IFU}$ abundances relative (O/H)$_{SLIT}$ abundances is less than the scatter
in abundance differences for any (except one) metallicity interval of 0.05 dex. Therefore it is not clear whether this weak correlation is meaningfull.

We find that the same calibration  can produce a close estimations of the abundances using the IFU spectra (of extracted H\,{\sc ii} regions or spatial samplings) from
the IFU measurements carried out with different spatial resolution and different native spatial samplings. This is in line with the finding in \citet{Mannucci2021} that the contribution of
the diffuse ionized gas to the large aperture  spectra of H\,{\sc ii} regions  has a secondary effect.

\section*{Acknowledgements}
We are grateful to the referee for his/her constructive comments. \\
We thank the VENGA collaboration for providing access to the data. \\ 
L.S.P acknowledges support from the Research Council of
Lithuania (LMTLT) (grant no. P-LU-PAR-22-7).                            \\
MALL acknowledges support from the Spanish grant PID2021-123417OB-I00.
SDP is grateful to the Fonds de Recherche du Québec - Nature et
Technologies. JVM and SDP acknowledge financial support from the Spanish
Ministerio de Econom\'ia y Competitividad under grants
AYA2016-79724-C4-4-P and PID2019-107408GB-C44, from Junta de Andaluc\'ia
Excellence Project P18-FR-2664, and also acknowledge support from the
State Agency for Research of the Spanish MCIU through the `Center of
Excellence Severo Ochoa' award for the Instituto de Astrof\'isica de
Andaluc\'ia (SEV-2017-0709).                                                   \\
This study makes use of the results based on the Calar Alto Legacy
Integral Field Area (CALIFA) survey (http://califa.caha.es/). \\

\appendix

\section{Abundances based on slit and IFU spectra in nearby galaxies}

Here we estimate the oxygen abundances based on the slit and the IFU spectra in seven nearby galaxies with available slit and IFU spectroscopy.
The (O/H)$_{R,IFU}$ and/or (O/H)$_{S,IFU}$ abundances for individual IFU spectra as well as the median abundances in bins of 0.1 in fractional radius $R/R_{25}$ are determined.
The median values of the slit spectra-based abundances (O/H)$_{SLIT}$ are determined using both the (O/H)$_{R}$ and the (O/H)$_{S}$ abundances for individual  H\,{\sc ii} regions together.
The medial value is determined if the number of points in the bin is larger than 3.
The abundances in each  galaxy are shown below,  Fig.~\ref{figure:ngc1058-r-oh} -- Fig.~\ref{figure:ngc6946-r-oh}.
The data for that sample of galaxies (together with the data for NGC~628) are presented in Fig.~\ref{figure:ohslit-ohifu} and Fig.~\ref{figure:gist-ohslit-ohifu} and
are used in discussion.

\subsection{NGC~1058}

\begin{figure}
\resizebox{1.00\hsize}{!}{\includegraphics[angle=000]{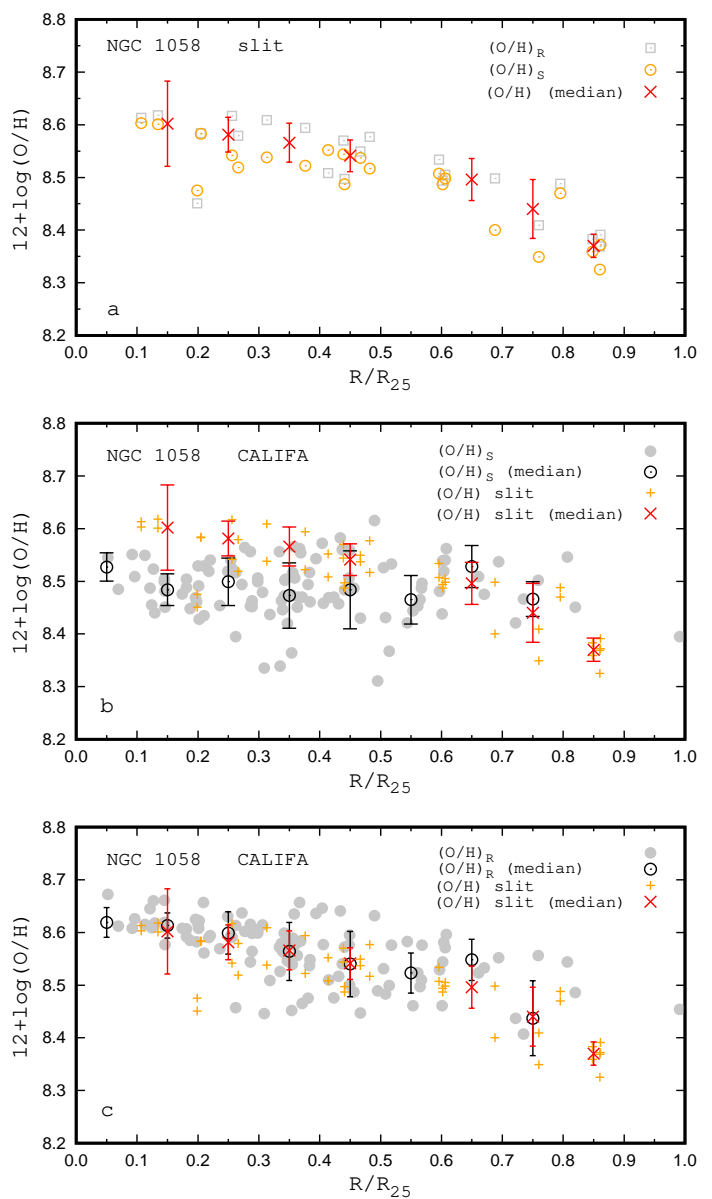}}
\caption{
  Radial oxygen abundance distribution in  NGC~1058.
  {\sl Panel {\bf a}:} oxygen abundances based on the slit spectra of H\,{\sc ii} regions. The squares denote the abundances estimated through the R calibration for individual
  measurements, the circles designate the (O/H)$_{S}$ abundances. The crosses mark the median values of the slit spectra-based abundances in bins of 0.1 in the fractional radius
  determined using both the (O/H)$_{R}$ and the (O/H)$_{S}$ abundances for individual  H\,{\sc ii} regions together.The bar shows the scatter in each bin.
  {\sl Panel {\bf b}:} grey points denote the (O/H)$_{S,IFU}$ abundances based on the extracted IFU (CALIFA) spectra of individual H\,{\sc ii} regions.
  The dark circles mark the median values in bins,  and bar shows the scatter in each bin.
  The crosses mark the median values of the slit spectra-based abundances and come from panel a.
  {\sl Panel {\bf c}:} the same as panel a but for the (O/H)$_{R,IFU}$ abundances.
}
\label{figure:ngc1058-r-oh}
\end{figure}

NGC~1058 is a Sc galaxy (morphological type code T = 5.1$\pm$0.9). The inclination angle of  NGC~1058 is $i$ = 15$\degr$, the position angle of the major axis PA = 145$\degr$
\citep{GarciaGomez2004}. The optical radius is 1.51 arcmin, or 90.6 arcsec \citep{RC3}. At the distance of $d$ = 10.6 Mpc \citep{Schmidt1994}, the physical optical radius is $R_{25}$ = 4.66 kpc.
The stellar mass (rescaled to the adopted distance) is M$_{\star}$ = 2.51 $\times$ 10$^{9}$ M$_{\sun}$ \citep{Leroy2019}.  

The slit spectra of H\,{\sc ii} regions in  NGC~1058 were measured by \citet{Ferguson1998} and \citet{Bresolin2019}.   The squares in panel a of Fig.~\ref{figure:ngc1058-r-oh} show
the oxygen abundances estimated through the R calibration in those H\,{\sc ii} regions. The circles in panel a of Fig.~\ref{figure:ngc1058-r-oh} are the oxygen abundances in those
H\,{\sc ii} regions estimated through the S calibration.
The median values in bins of 0.1 in fractional radius $R/R_{25}$ determined using both the (O/H)$_{R}$ and the (O/H)$_{S}$ abundances for individual  H\,{\sc ii} regions together
are denoted by crosses.  The bar shows the scatter in each bin.

The IFU spectroscopy of NGC~1058 is carried out within the CALIFA survey. The grey points in panel b of Fig.~\ref{figure:ngc1058-r-oh} show the IFU spectra-based (O/H)$_{S}$ abundances  of
H\,{\sc ii} regions as a function of radius in  NGC~1058. The line flux measurements in the spectra of H\,{\sc ii} regions are taken from the catalog of  H\,{\sc ii}
regions (cited above). The median values in bins of 0.1 in fractional radius $R/R_{25}$ are denoted by dark circles.  The bar shows the scatter in each bin.
Panel c of Fig.~\ref{figure:ngc1058-r-oh} shows the same as panel b but for (O/H)$_{R22}$ abundances.

\subsection{NGC~1672}

\begin{figure}
\resizebox{1.00\hsize}{!}{\includegraphics[angle=000]{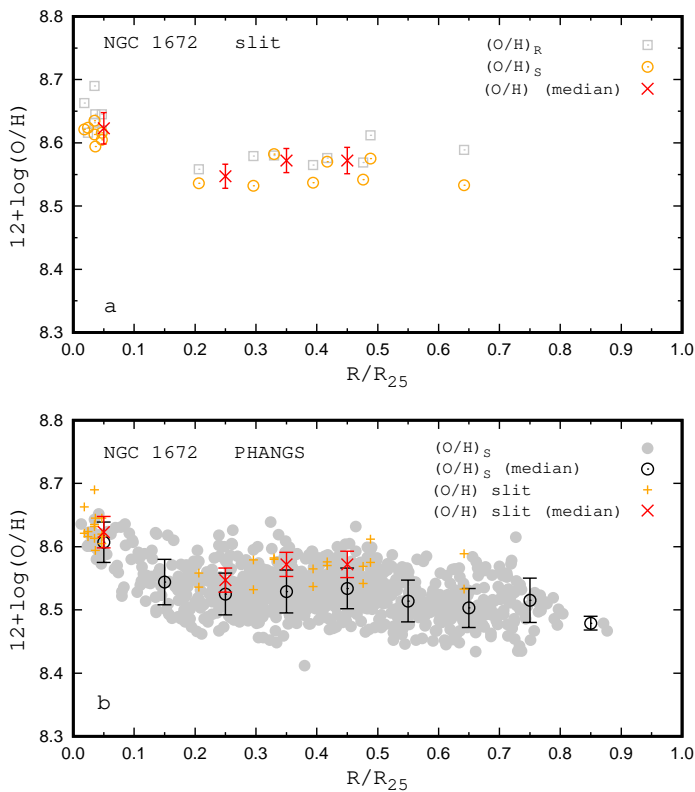}}
\caption{
  Radial oxygen abundance distribution in  NGC~1672.
  {\sl Panel {\bf a}:} oxygen abundances based on the slit spectra of H\,{\sc ii} regions. 
  {\sl Panel {\bf b}:} oxygen abundances based on the IFU (PHANGS) spectra of H\,{\sc ii} regions.
  The notations are the same as in Fig.~\ref{figure:ngc1058-r-oh}. 
}
\label{figure:ngc1672-r-oh}
\end{figure}

NGC~1672 is a Sb galaxy  (morphological type code T = 3.3$\pm$0.6). The inclination angle of  NGC~1672 is $i$ = 43$\degr$, and the position angle of the major axis
PA = 134$\degr$ \citep{Lang2020}. The optical radius is 3.30 arcmin or 198.2 arcsec \citep{RC3}. At the distance of $d$ = 19.40 Mpc \citep{Anand2021},
the physical optical radius is $R_{25}$ = 18.64 kpc. The stellar mass is M$_{\star}$ = 5.37 $\times$ 10$^{10}$ M$_{\sun}$  \citep{Leroy2021}. NGC~1672 hosts a central black
hole of  mass logM$_{BH}$ = 7.08$\pm 0.90$ in solar mass \citep{Davis2014}.

The slit spectra of H\,{\sc ii} regions in  NGC~1672 were measured by \citet{Storchi1996}. The estimated (O/H)$_{R}$ and (O/H)$_{S}$ abundances in individual H\,{\sc ii} HII regions
are shown in panel a of Fig.~\ref{figure:ngc1672-r-oh}  by squares and circles, respectively. The crosses are the median values in bins for those data. 
The IFU spectroscopy of  NGC~1672 were carried out within the framework of the PHANGS programme \citep{Kreckel2019}. The catalog of the H II regions in  NGC~1672 was created
on the base of those measurements. Unfortunately only the red spectra  over the wavelength range covering 4800-9300{\AA} were obtained. Since the line [O\,{\sc ii}]$\lambda\lambda$3727,3729
is not measured then the R calibration cannot be applied to those  H\,{\sc ii} regions.  The grey circles in panel b of Fig.~\ref{figure:ngc1672-r-oh} denote  the oxygen abundances
in individual H\,{\sc ii} regions estimated through the S calibration, and the dark circles are the median values in bins for those data.

\subsection{NGC~2835}

\begin{figure}
\resizebox{1.00\hsize}{!}{\includegraphics[angle=000]{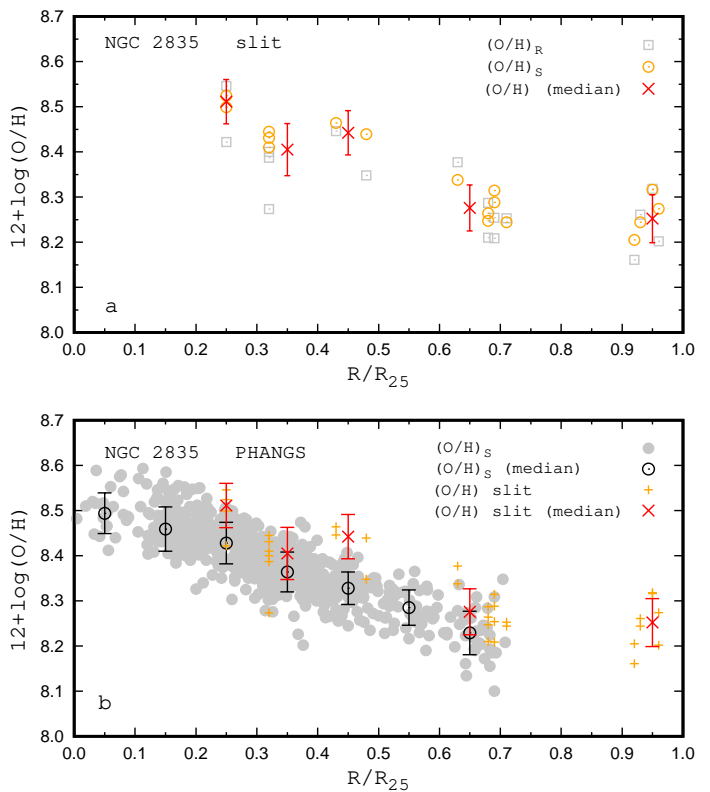}}
\caption{
  Radial oxygen abundance distribution in  NGC~2835.
  {\sl Panel {\bf a}:} oxygen abundances based on the slit spectra of H\,{\sc ii} regions. 
  {\sl Panel {\bf b}:} oxygen abundances based on the IFU (PHANGS) spectra of H\,{\sc ii} regions.
  The notations are the same as in Fig.~\ref{figure:ngc1058-r-oh}. 
}
\label{figure:ngc2835-r-oh}
\end{figure}

NGC~2835 is a Sc galaxy  (morphological type code T = 5.0$\pm$0.4). The inclination angle of  NGC~2835 is $i$ = 51$\degr$ \citep{Ryder1995}, the position angle of the major axis
PA = 1$\degr$ \citep{Lang2020}. The optical radius is 3.30 arcmin or 198.2 arcsec \citep{RC3}. At the distance of $d$ = 12.22 Mpc \citep{Anand2021}, the physical
optical radius is $R_{25}$ = 11.74 kpc. The stellar mass is M$_{\star}$ = 1.0 $\times$ 10$^{10}$ M$_{\sun}$ \citep{Leroy2021}. The mass of the central black hole in  NGC~2835 is logM$_{BH}$ =
6.72$\pm 0.30$ in solar mass \citep{Davis2014}.
The slit spectra of H\,{\sc ii} regions in  NGC~2835 were measured by \citet{Ryder1995}. Panel a of Fig.~\ref{figure:ngc2835-r-oh} shows the oxygen abundances  in those H\,{\sc ii} regions:
the squares mark the (O/H)$_{R}$ abundances and the circles denote the (O/H)$_{S}$ abundances. The crosses are the median values in bins of 0.1 in fractional radius $R/R_{25}$. 
The IFU spectroscopy of  NGC~2835 were carried out within the framework of the PHANGS programme and a catalog of H\,{\sc ii} regions was created \citep{Kreckel2019}.  The grey
circles in panel b of Fig.~\ref{figure:ngc2835-r-oh} denote  the oxygen abundances in individual H\,{\sc ii} regions estimated through the S calibration,
and the dark circles are the median values in bins for those data.

\subsection{NGC~2903}

\begin{figure}
\resizebox{1.00\hsize}{!}{\includegraphics[angle=000]{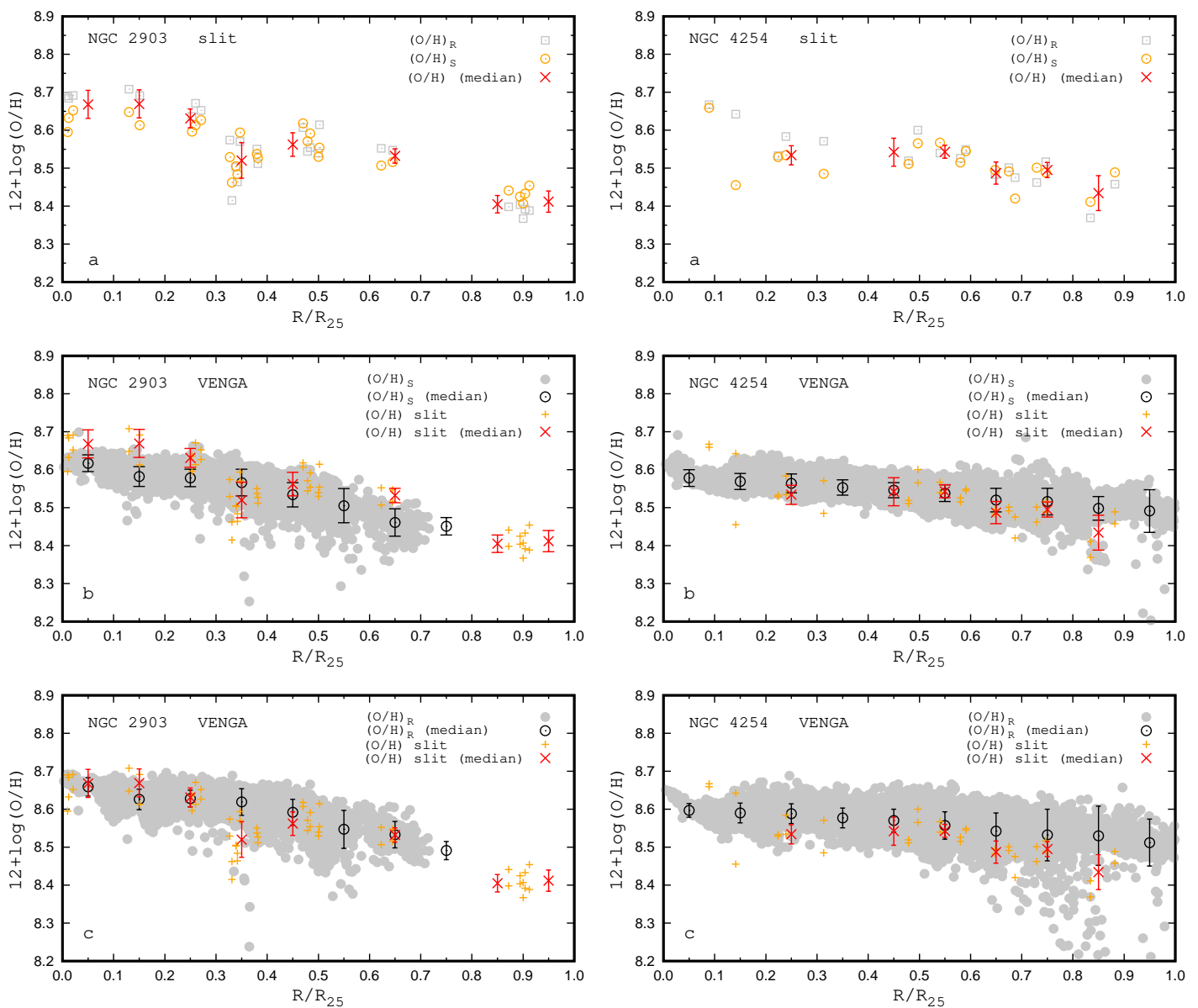}}
\caption{
  Radial oxygen abundance distribution in  NGC~2903.
  {\sl Panel {\bf a}:} oxygen abundances based on the slit spectra of H\,{\sc ii} regions.
  {\sl Panel {\bf b}:} oxygen abundances estimated through the S calibration from the IFU (VENGA) spectra of fibres.
  {\sl Panel {\bf c}:} oxygen abundances obtained through the R calibration from the IFU (VENGA) spectra of fibres.
  The notations are the same as in Fig.~\ref{figure:ngc1058-r-oh}. 
}
\label{figure:ngc2903-r-oh}
\end{figure}

The nearby galaxy NGC~2903 is a Sbc spiral galaxy (morphological type code T = 4.0$\pm$0.1). Its inclination angle is $i$ = 65$\degr$ and the position angle of the major axis
PA = 204$\degr$ \citep{deBlok2008}. The optical radius of  NGC~2903 is R$_{25}$ = 5.87 arcmin \citep{Walter2008}. The distance to  NGC~2903 is $d$ = 8.9 Mpc 
\citep{Drozdovsky2000} that results in the physical optical radius of $R_{25}$ = 15.21 kpc. The mean value of the estimations of the stellar mass of  NGC~2903 (rescaled to
the adopted distance) is M$_{\star}$ = 3.33 $\times$ 10$^{10}$ M$_{\sun}$ \citep{Jarrett2019,Leroy2021}.  NGC~2903 hosts a supermassive black hole of mass logM$_{BH}$ = 7.06$^{+0.28}_{-7.06}$
in solar mass \citep{vandenbosch2016}.

The slit spectra of H\,{\sc ii} regions in the disk of NGC~2903 were measured by \citet{McCall1985},  \citet{vanZee1998}, \citet{Bresolin2005}, and \citet{Diaz2007}.
The squares in panel a of Fig.~\ref{figure:ngc2903-r-oh} show the oxygen abundances estimated through the R calibration in those H\,{\sc ii} regions, and the circles  denote the oxygen
abundances estimated through the S calibration. The crosses are the median values in bins of 0.1 in fractional radius $R/R_{25}$. 
The IFU (fibre) spectroscopy of  NGC~2903 was performed within the framework of the VENGA survey \citep{Blanc2013,Kaplan2016}. The grey points in panel b of Fig.~\ref{figure:ngc2903-r-oh} show the
IFU spectra-based (O/H)$_{S,IFU}$ abundances in the individual fibres, and the dark circles are the median values in bins for those data.
Panel c of Fig.~\ref{figure:ngc2903-r-oh} shows the same as panel b but for (O/H)$_{R,IFU}$ abundances.

\subsection{NGC~4254}

\begin{figure}
\resizebox{1.00\hsize}{!}{\includegraphics[angle=000]{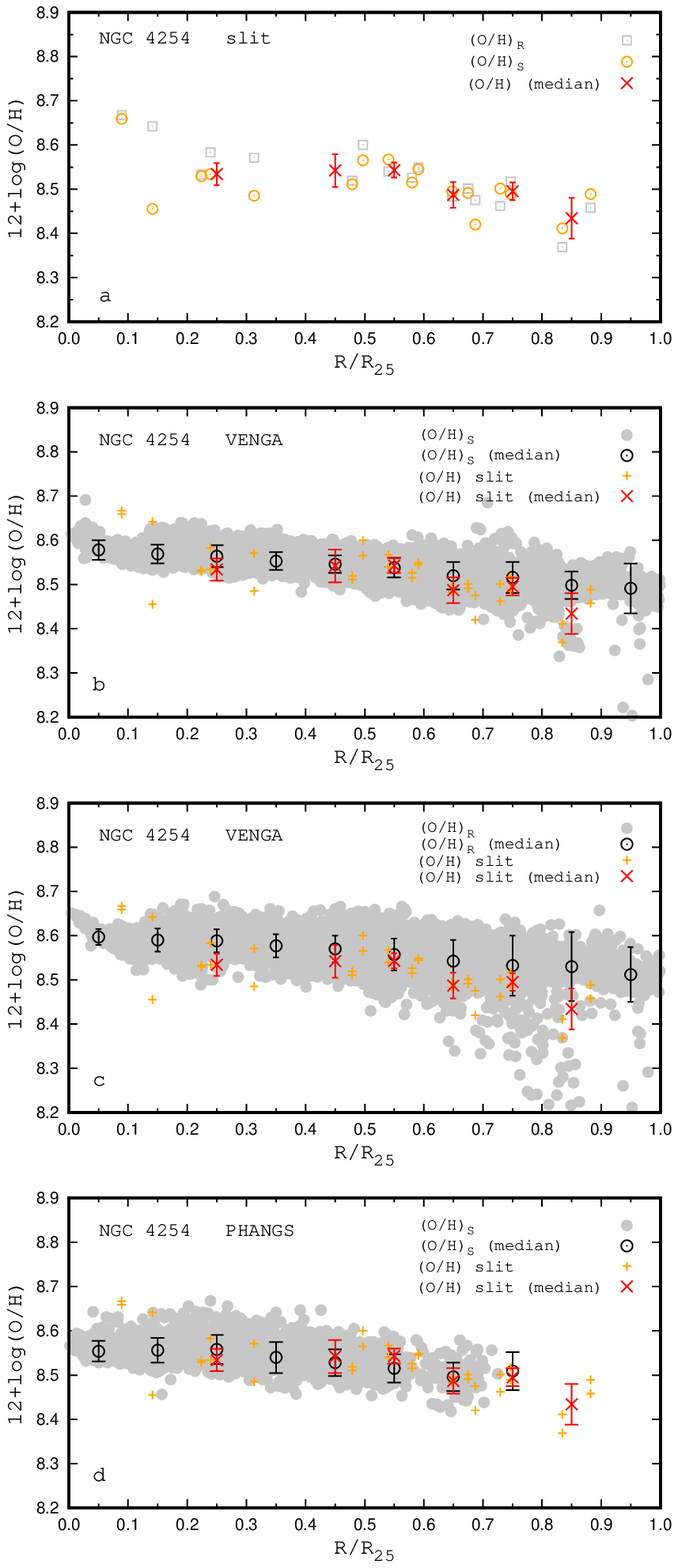}}
\caption{
  Radial oxygen abundance distribution in  NGC~4254.
  {\sl Panel {\bf a}:} oxygen abundances based on the slit spectra of H\,{\sc ii} regions.
  {\sl Panel {\bf b}:} oxygen abundances estimated through the S calibration from the IFU (VENGA) spectra of fibres.
  {\sl Panel {\bf c}:} oxygen abundances obtained through the R calibration from the IFU (VENGA) spectra of fibres.
  {\sl Panel {\bf d}:} oxygen abundances  estimated through the S calibration from  the IFU (PHANGS) spectra of H\,{\sc ii} regions.
  The notations are the same as in Fig.~\ref{figure:ngc1058-r-oh}. 
}
\label{figure:ngc4254-r-oh}
\end{figure}

NGC~4254 (M~99) is a bright Sc galaxy  (morphological type code T = 5.2$\pm$0.7) in the Virgo Cluster. The inclination angle of  NGC~4254 is $i$ = 34$\degr$, the position
angle of the major axis PA = 68$\degr$ \citep{Lang2020}. The optical radius is 2.68 arcmin \citep{RC3}. At the distance of $d$ = 13.1 Mpc \citep{Anand2021},
the physical optical radius is $R_{25}$ = 10.23 kpc. The stellar mass is M$_{\star}$ = 2.63 $\times$ 10$^{10}$ M$_{\sun}$ \citep{Leroy2021}.  

The slit spectra of H\,{\sc ii} regions in the disk of NGC~4254 were measured by \citet{McCall1985},  \citet{Shields1991}, and \citet{Henry1994}.  The squares in panel a of
Fig.~\ref{figure:ngc4254-r-oh} show the oxygen abundances estimated through the R calibration in those H\,{\sc ii} regions. 
The circles are the oxygen abundances  estimated through the S calibration. The crosses are the median values in bins of 0.1 in fractional radius $R/R_{25}$. 
The IFU spectroscopy of  NGC~4254 was carried out within the framework of the VENGA survey \citep{Blanc2013,Kaplan2016}. The grey points in panel b of Fig.~\ref{figure:ngc4254-r-oh} show
the individual fibre (O/H)$_{S,IFU}$ abundances, and the dark circles are the median values in bins for those data.
Panel c of Fig.~\ref{figure:ngc4254-r-oh} shows the same as panel b but for the (O/H)$_{R,IFU}$ abundances. 
The IFU spectroscopy of  NGC~4254 was also carried out within the framework of the PHANGS programme and the catalog of  H\,{\sc ii} regions is constructed \citep{Kreckel2019}.
The grey points in panel d of Fig.~\ref{figure:ngc4254-r-oh} show the (O/H)$_{S,IFU}$ abundances based on the IFU spectra of H\,{\sc ii} regions,
and the dark circles are the median values in bins for those data.

\subsection{NGC~5194}

\begin{figure}
\resizebox{1.00\hsize}{!}{\includegraphics[angle=000]{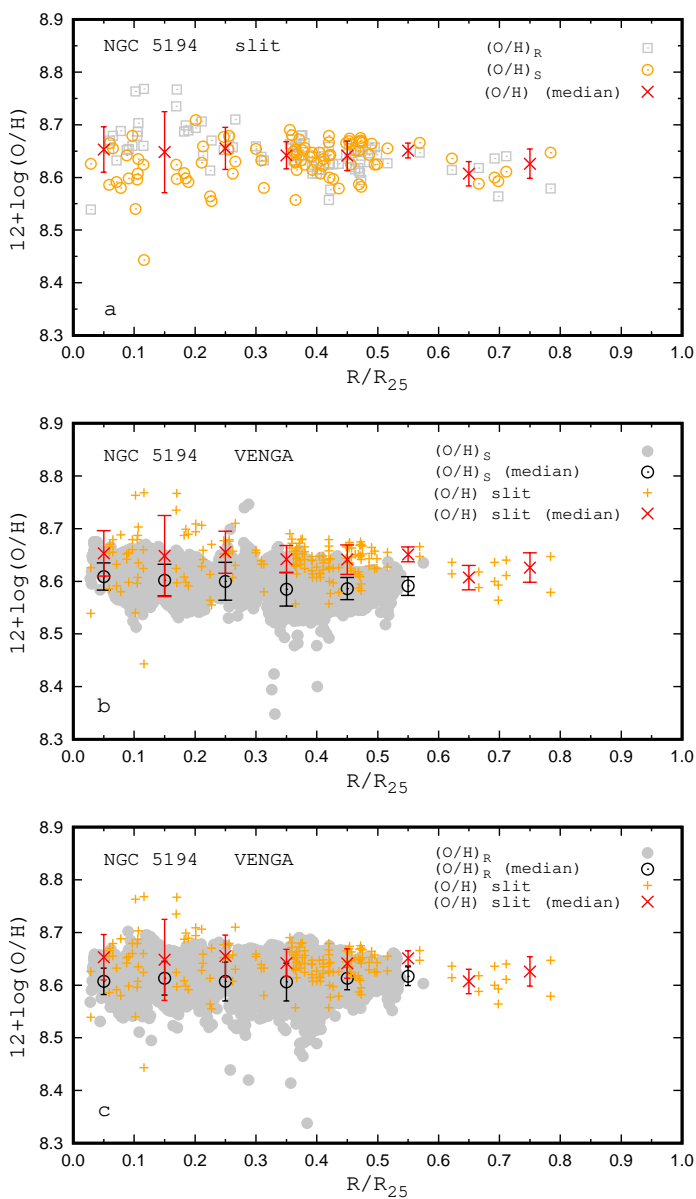}}
\caption{
  Radial oxygen abundance distribution in  NGC~5194.
  {\sl Panel {\bf a}:} oxygen abundances based on the slit spectra of H\,{\sc ii} regions.
  {\sl Panel {\bf b}:} oxygen abundances estimated through the S calibration from the IFU (VENGA) spectra of fibres.
  {\sl Panel {\bf c}:} oxygen abundances obtained through the R calibration from the IFU (VENGA) spectra of fibres.
  The notations are the same as in Fig.~\ref{figure:ngc1058-r-oh}. 
}
\label{figure:ngc5194-r-oh}
\end{figure}

The nearby galaxy NGC~5194 (M~51a, the Whirlpool Galaxy) is a SABb spiral galaxy (morphological type code T = 4.0$\pm$0.3). The bright disk of NGC~5194 ends abruptly at about 5 arcmin
radius in both the optical images and the  H\,{\sc i}. The velocity structure of the gas in NGC~5194 is extremely complicated and difficult to interpret \citep{Rots1990}. There is
the misalignment of the major axes of the H\,{\sc i} distribution and the velocity field. Therefore the geometrical parameters of  NGC~5194 are rather uncertain. \citet{Tamburro2008}
derive the following geometrical projection parameters of the NGC~5194 galactic disk: position angle PA = 172$\degr$ and inclination $i$ = 42$\degr$. \citet{Colombo2014} undertake a detailed
kinematic study of NGC~5194 and found a position angle PA = (173 $\pm$ 3)$\degr$, and an inclination $i$ = (22 $\pm$ 5)$\degr$. The geometrical parameters of  NGC~5194 obtained by \citet{Colombo2014}
are used here. We adopt the optical radius of NGC~5194 to be $R_{25}$ = 5.61 arcmin \citep{RC3}. It should be noted that the value of the optical radius of R$_{25}$ = 3.88 arcmin is used
for NGC~5194 within The  H\,{\sc i} Nearby Galaxy Survey (THINGS) \citep{Walter2008}. There are recent distance estimations for  NGC~5194 through the tip of the red giant branch (TRGB)
method based on the Hubble Space Telescope measurements. We adopt here the distance to  NGC~5194 as $d$ = 8.58  Mpc,  obtained by \citet{McQuinn2017}. The optical radius of  NGC~5194 is
$R_{25}$ = 14.00 kpc with adopted distance. The stellar mass rescaled to the adopted distance is M$_{\star}$ = 4.54 $\times$ 10$^{10}$ M$_{\sun}$ \citep{Leroy2008,Jarrett2019}. 

The slit spectra of H\,{\sc ii} regions in  NGC~5194 were measured in a number of works \citep{McCall1985,Diaz1991,Bresolin1999,Bresolin2004,Garnett2004,Croxall2015}. Panel
a of Fig.~\ref{figure:ngc5194-r-oh} shows the oxygen abundances estimated through the R calibration (squares) and through the S calibration (circles) in those H\,{\sc ii} regions. 
The crosses are the median values in bins of 0.1 in fractional radius $R/R_{25}$. The IFU spectroscopy of  NGC~5194 was performed within the framework of the VENGA survey
\citep{Blanc2013,Kaplan2016}. The grey points in panel b of Fig.~\ref{figure:ngc5194-r-oh} show the fibre (O/H)$_{S}$ abundances, and the dark circles are the median values in bins.

\subsection{NGC~6946}

\begin{figure}
\resizebox{1.00\hsize}{!}{\includegraphics[angle=000]{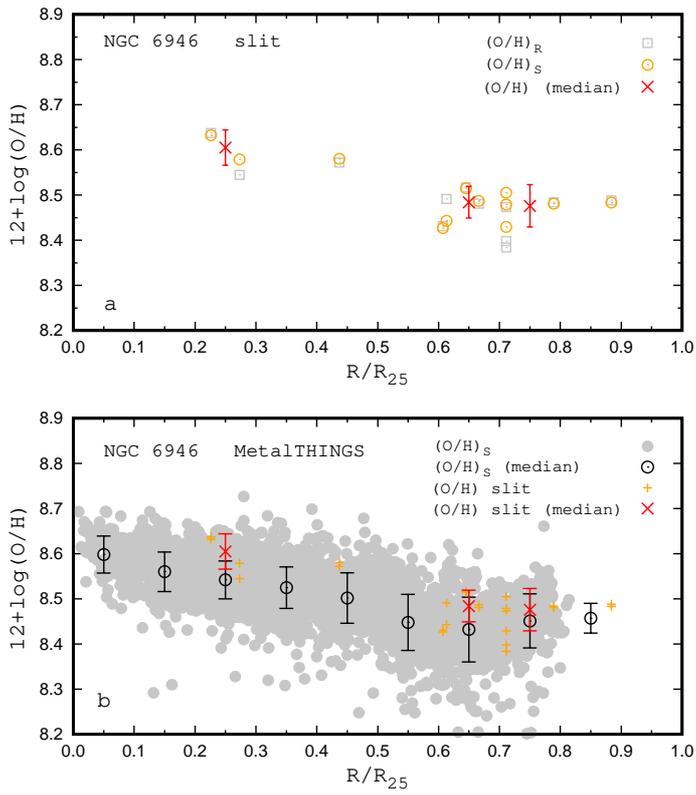}}
\caption{
  Radial oxygen abundance distribution in  NGC~6946.
  {\sl Panel {\bf a}:} oxygen abundances based on the slit spectra of H\,{\sc ii} regions.
  {\sl Panel {\bf b}:} oxygen abundances estimated through the S calibration from the IFU (Metal-THINGS) spectra of fibres.
  The notations are the same as in Fig.~\ref{figure:ngc1058-r-oh}. 
}
\label{figure:ngc6946-r-oh}
\end{figure}

NGC~6946 is SABc galaxy (morphological type code T = 5.9$\pm$0.3). The inclination angle of  NGC~6946 is $i$ = 33$\degr$, and the position angle of the major axis PA = 243$\degr$
\citep{deBlok2008}. The optical radius is 5.74 arcmin \citep{RC3}. At the distance  of $d$ = 7.34 Mpc \citep{Anand2021}, the physical optical radius is $R_{25}$ = 12.26
kpc. The stellar mass (mean value) is M$_{\star}$ = 2.85 $\times$ 10$^{10}$ M$_{\sun}$ \citep{Jarrett2019,Leroy2019}.  

The slit spectra of H\,{\sc ii} regions in  NGC~6946 were measured by \citet{McCall1985}, \citet{Ferguson1998}, and \citet{GarciaBenito2010}. The squares in panel a of
Fig.~\ref{figure:ngc6946-r-oh} show the oxygen abundances estimated through the R calibration (squares) and through the S calibration (circles) in those H\,{\sc ii} regions.
The crosses are the median values in bins of 0.1 in fractional radius $R/R_{25}$. The IFU spectroscopy of  NGC~6946 were carried out within the framework of the Metal-THINGS programme
\citep{LaraLopez2022}. Only the red spectra  over the wavelength range covering 4800-9300{\AA} were obtained.  The grey circles in panel b of Fig.~\ref{figure:ngc6946-r-oh} denote
the oxygen abundances in individual fibres estimated through the S calibration, and the dark circles are the median values in bins.

\end{document}